\documentclass[12pt]{emulateapj}

\newcommand{\xmmnewton}{{\it XMM-Newton}}
\newcommand{\rosat}{{\it ROSAT}}

\newcommand{\einstein}{{\it Einstein}}
\newcommand{\chandra}{{\it Chandra}}

\newcommand{\hst}{{\it HST}}

\newcommand{\spitzer}{{\it Spitzer}}
\newcommand{\etal}{et al.}

\newcommand{\nh}{\mbox {$N_{\rm H}$}}

\newcommand{\hii}{H\,{\sc ii}}

\newcommand{\about}{$\sim$\kern.03em}
\newcommand{\ergs}[1]{$\times 10^{#1}$ erg s$^{-1}$}

\newcommand{\chase}{ChASeM33}
\newcommand{\m}{M\,33}
\newcommand{\halpha}{${\rm H}\alpha$}

\shorttitle{\chase}
\shortauthors{Plucinsky et al.}

\begin{document}


\title{\chandra\/ ACIS Survey of \m\ (\chase): A First Look}


\author{Paul P. Plucinsky\altaffilmark{1},
Benjamin Williams\altaffilmark{2},
Knox S. Long\altaffilmark{3},
Terrance J. Gaetz\altaffilmark{1},
Manami Sasaki\altaffilmark{1},
Wolfgang Pietsch\altaffilmark{4},
Ralph T\"ullmann\altaffilmark{1},
Randall K. Smith\altaffilmark{5,6}, 
William P. Blair\altaffilmark{5},
David Helfand\altaffilmark{7},
John P. Hughes\altaffilmark{8}, 
P. Frank Winkler\altaffilmark{9},
Miguel de Avillez\altaffilmark{10},
Luciana Bianchi\altaffilmark{5},
Dieter Breitschwerdt\altaffilmark{11},
Richard J. Edgar\altaffilmark{1}, 
Parviz Ghavamian\altaffilmark{5},
Jonathan Grindlay\altaffilmark{1}, 
Frank Haberl\altaffilmark{4},
Robert Kirshner\altaffilmark{1},
Kip Kuntz\altaffilmark{5,6},
Tsevi Mazeh\altaffilmark{12},
Thomas G. Pannuti\altaffilmark{13},
Avi Shporer\altaffilmark{12}, and
David A. Thilker\altaffilmark{5}
}

\altaffiltext{1}{Harvard-Smithsonian Center for Astrophysics,
    60 Garden Street, Cambridge, MA 02138;
plucinsky@cfa.harvard.edu}

\altaffiltext{2}{Astronomy Department, University of Washington, Seattle, 
WA 98195}

\altaffiltext{3}{Space Telescope Science Institute, 
3700 San Martin Drive, Baltimore, MD 21218}

\altaffiltext{4}{Max-Planck-Institut f\"ur Extraterrestrische Physik,
85741 Garching, Germany}

\altaffiltext{5}{Department of Physics and Astronomy, 
Johns Hopkins University, 3400 North Charles Street, Baltimore, MD 21218}

\altaffiltext{6}{NASA Goddard Space Flight Center, Code 662, Greenbelt, 
MD 20771} 

\altaffiltext{7}{Columbia Astrophysics Laboratory, 550 W. 
$120^{\mathrm{th}}$ St., New York, NY, 10027}

\altaffiltext{8}{Department of Physics and Astronomy, Rutgers University, 136
Frelinghuysen Road, Piscataway, NJ 08854-8019}

\altaffiltext{9}{Department of Physics, Middlebury College,  
Middlebury, VT 05753}

\altaffiltext{10}{Department of Mathematics, University of Evora, 
R. Romao Ramalho 59, 7000 Evora, Portugal}

\altaffiltext{11}{Institut f\"ur Astronomie, Universit\"at Wien,
T\"urkenschanzstr. 17, A-1180 Wien, Austria}

\altaffiltext{12}{School of Physics and Astronomy, Raymond and Beverly Sackler
Faculty of Exact Sciences, Tel Aviv University, Tel Aviv, Israel 69978}

\altaffiltext{13}{Space Science Center, 200A Chandler Place, Morehead State
University, Morehead, KY 40351}


\begin{abstract}

We present an overview of the \chandra\/ ACIS
Survey of \m\ (\chase): A Deep Survey of the Nearest Face-on Spiral
Galaxy.  The 1.4 Ms survey covers the galaxy out to $R \approx
18\arcmin\/ (\approx 4$\ kpc). These data provide the most intensive,
high spatial resolution assessment of the X-ray source populations
available for the confused inner regions of \m.
Mosaic images of the \chase\/
observations show several hundred individual X-ray sources as well as soft
diffuse emission from the hot interstellar medium.  Bright, extended
emission surrounds the nucleus and is also seen from the giant \hii\/
regions NGC\,604 and IC\,131.  Fainter extended emission and numerous
individual sources appear to trace the inner spiral structure.  The
initial source catalog, arising from $\sim$~$2/3$ of the expected
survey data, includes 394 sources significant at the
$3\sigma$ confidence level or greater, down to a limiting luminosity
(absorbed) of $\sim$1.6\ergs{35} (0.35 -- 8.0~keV). 
The hardness ratios of the sources 
separate those with soft, thermal spectra such as supernova remnants from
those with hard, non-thermal spectra such as X-ray binaries and 
background active galactic nuclei.  
Emission extended beyond the
\chandra\/ point spread function is evident in 23 of the 394
sources. Cross-correlation of the \chase\/ sources against previous
catalogs of X-ray sources in \m\/ results in matches for the vast 
majority of the brighter sources and shows 28 \chase\/ sources within 
10\arcsec\/ of supernova remnants identified by prior optical and
radio searches. This brings the total number of such associations to
31 out of 100 known supernova remnants in \m\/.

\end{abstract}




\section{Introduction}

\m\ is a late-type Sc spiral galaxy  located at a distance of 
817$\pm$58~kpc \citep{2001ApJ...553...47F} and is the third largest 
spiral in the Local Group after M\,31 and the Milky Way.
The galaxy's intermediate inclination angle of 56\degr$\pm$1\degr\ 
\citep{1989AJ.....97...97Z}, low foreground absorption
\citep[galactic $\nh\sim6.0\times10^{20}$~cm$^{-2}$,][]
{dickey1990,stark1992}, and 
modest internal extinction make it the ideal target for 
exploring the global structure of the interstellar medium (ISM) and the 
stellar populations that shape it.   

\m\ has been the subject of numerous previous X-ray studies,
beginning with the \einstein\ Observatory  \citep{long81, markert83, trinchieri88},
 which discovered 17 sources (including one that
is probably a foreground star and one a background active 
galactic nuclei (AGN)).
The most dominant source by far is associated with the nucleus, with 
$L_X \sim 1$\ergs{39}, making it the brightest persistent source
in the Local Group.  In addition to these individual sources, substantial
unresolved emission was detected.
 \rosat\ observations by \citet{schulman95} and \citet{long96} expanded
the list of individual X-ray sources to 57 and showed that diffuse emission 
may trace the spiral arms near the nucleus.  Through retrospective combination of
all the \rosat\ observations, \citet{haberl01} found 184 X-ray sources within
50\arcmin\/ of the nucleus, a larger radius than in previous studies and 
including a number of foreground and background objects.

Most recently, \citet{pietsch04} and \citet{misanovic06} 
reported 447 individual X-ray sources in a deep \xmmnewton\/ survey of
\m\ within the $D_{25}$\ optical elliptical isophote  
(semi-major axis $\sim 32\arcmin\/$).  
They characterized some of these sources based on X-ray hardness ratios and
identified several dozen of these with supernova remnants, X-ray
binaries and  supersoft sources in \m\, 
based on cross-correlations with optical, radio, and infrared catalogs.
In addition, they identified a number of foreground
stars and background galaxies.  Two of the transient sources were
identified with optical novae in \m\/ \citep{pietsch2005} and other 
transients were identified as probable X-ray binaries 
\citep{misanovic06}.  \cite{Grimm05} constructed a source list
containing 261 sources and a
luminosity function from \chandra\/ observations that covered a 
$0.16~{\rm deg^2}$ region including the center of the galaxy and a
region to the northwest of the center.  

It is apparent in X-ray maps from \xmmnewton\/ and earlier missions that 
\m\ is very confused in its inner regions where the density of individual 
sources becomes highest and the truly diffuse emission is strongest.
We therefore proposed to use \chandra's superior angular resolution 
with the Advanced CCD Imaging Spectrometer 
\citep[ACIS,][]{2003SPIE.4851...28G} to conduct the \chandra\/ 
ACIS Survey of M33 (ChASeM33).
We are obtaining deep, high-resolution images of the inner regions of 
\m\, out to a radius of $\approx 18\arcmin\/$ or $\approx 4$\ kpc
(1\arcsec $\sim$ 4~pc).
A total of 1.4~Ms of \chandra\/ observations
are devoted to this project, divided among seven ACIS  fields.
Each field is observed for a total of 200~ks divided between two 
(or more) visits
separated by 5--12 months.  The first observation of the survey was
executed in September 2005. In this paper, we report on the first
group of these observations in which each field was observed at least
once.  This first group of observations allows us to create mosaic
images of the region covered by the survey and to produce an initial source
catalog, at typically half the depth of the complete survey.
The primary objectives of the \chase\/ are to:
(1) establish the morphology of extended sources such as SNRs and
superbubbles, 
(2) locate the optical counterparts -- globular clusters, stars, 
and background AGN -- of a large fraction of the point sources, and 
(3) study the association of various sources with the Population I and II
components of \m\/. 
In addition, \chase\/ will provide long-duration 
observations for variability studies and transient detection. 
Given that much work can be pursued on these objectives with the partial
survey data, we present here an overview paper that
describes the survey and its advantages over previous surveys, 
discusses analysis techniques, and presents an
intermediate source list with basic source characterization. 

Some individual results from the \chase\/ survey have already been published.
Early observations of a field containing the  eclipsing X-ray binary 
(XRB) \m\ 
X-7  resolved the eclipse ingress and egress for 
the first time and indicated that 
\m\ X-7 is the first known eclipsing black hole  X-ray 
binary \citep{pietsch06a}.  
Furthermore, we found the eclipse and orbital period 
of the \xmmnewton\/ source \#47 from \citet[][hereafter PMH04]{pietsch04} in 
the \chase\/ observations \citep{pietsch06b}. 
The likely optical counterpart also shows variability; therefore, this source 
is the second high-mass X-ray binary detected in \m\/.  
Finally, the \chase\/ data for the brightest X-ray SNR in \m\ 
\citep[source \#21 from the SNR catalog of][hereafter GKL98]
{Gordon98}  are able to resolve the SNR from the 
giant \hii\/  region NGC~592.  Detailed spatial and spectral analysis 
of X-ray and other data shows a complex multiwavelength morphology
indicating a shell expanding into an inhomogeneous medium, typical of an
active star-forming region and suggesting that the SNR is 
indeed embedded within the \hii\/ region \citep{gaetz07}.

The organization of this paper is as follows.  In \S~\ref{sec-obs} we describe 
the observations in detail (\S~\ref{sec-data}), 
the initial processing of the data (\S~\ref{sec-proc}) , 
the creation of the mosaic images (\S~\ref{sec-mosaics}),
and the creation of the source catalog (\S~\ref{sec-catalog}). 
In \S~\ref{sec-analysis}, we describe 
the characterization of the sources based on hardness ratios
(\S~\ref{sec-HRs}), 
the identifications of the sources based on other catalogs
(\S~\ref{sec-IDs}),
the bright \hii\/ region NGC~604 (\S~\ref{sec-NGC604}), and  
the complicated southern arm region (\S~\ref{sec-southarm}).
Finally, in \S~\ref{sec-summary}, we summarize the results of the
paper.

\section{Observations and Data Reduction}\label{sec-obs}

\subsection{Observations}\label{sec-data}

\begin{figure}
\centering
\plotone{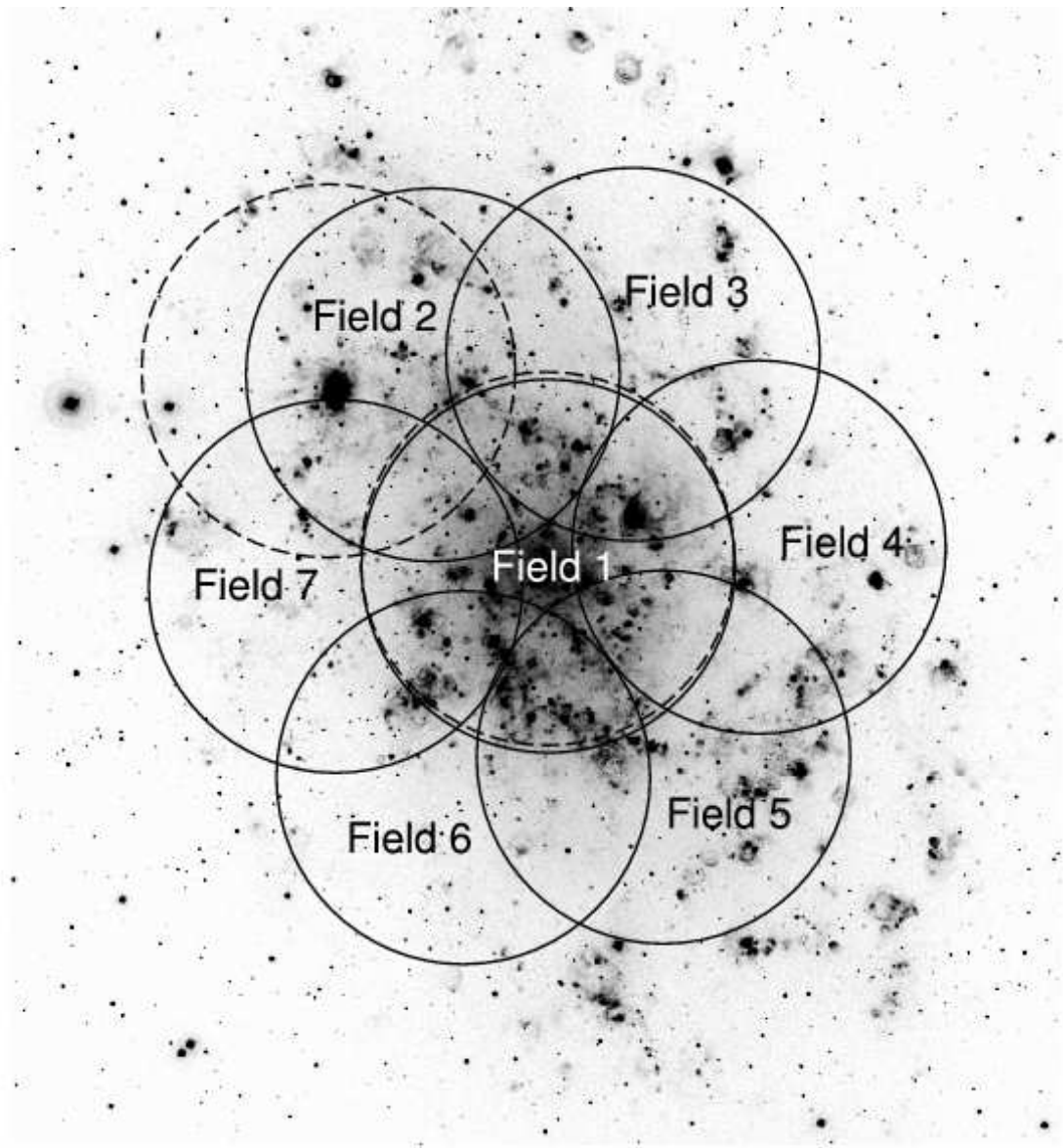}
\caption{\label{fovs}
Field of views of \chase\/ pointings are plotted as 8\arcmin\/ circles
(solid lines) on an \halpha\/ image from the Burrell Schmidt telescope
on Kitt Peak \citep{2006AAS...39....80M}.
The ObsIDs 1730 and 2023 are shown as the circles with dashed lines.
}
\end{figure}

The \chase\/ survey is designed to cover the central region of \m\/ at 
a resolution $\lesssim 5\arcsec\/$ in order to resolve the source confusion 
in the inner galaxy found in earlier surveys, and to increase the
number of known associations between X-ray sources and objects
identified at other wavelengths. 
We chose seven overlapping ACIS-I fields, arranged 
such that all regions within a galactocentric radius of $\sim 4$\ kpc 
lie no farther than 8\arcmin\/ off-axis in one or more fields  
(at 8\arcmin\/ off-axis, the 50\% (90\%) encircled-energy radius at
1.5~keV of the 
\chandra\/ PSF is 3\arcsec\/ (7\arcsec\/)).  Figure~\ref{fovs} shows the 
positions of the \chase\/ fields on a deep \halpha\/ image taken from 
the 0.6m Burrell Schmidt telescope on Kitt Peak  
\citep{2006AAS...39....80M}.  We chose the ACIS-I array for the survey 
because of its larger field of view (FOV) at small off-axis angles
compared to the ACIS-S array.  The circles drawn in Figure~\ref{fovs}
are contained inside the square $17\arcmin\times17\arcmin$ FOVs of the
ACIS-I array and indicate the regions within 8\arcmin\/ of the on-axis
point.  

\begin{figure}
\centering
\plotone{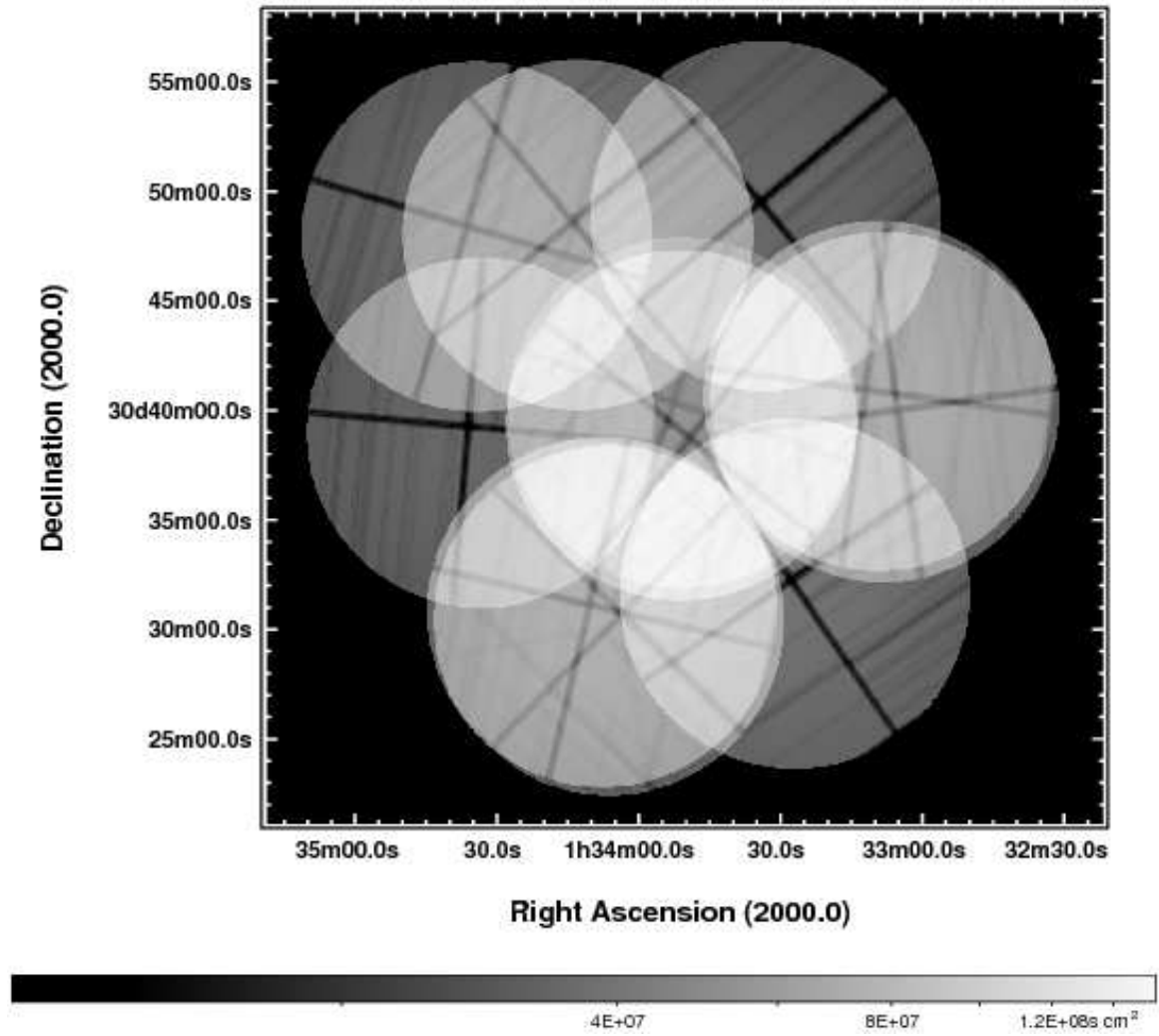}
\caption{\label{expmap}
Exposure map of all \m\ pointings with ACIS-I as the primary
instrument (the \chase\/ observations and OBSIDs 1730 and 2023) in
units of ${\rm s~cm^2}$ , only
including data with off-axis angles $<$ 8\arcmin. 
To compute the exposure map, spectral-weighted instrument maps were
used as discussed in \S~\ref{sec-mosaics}.
}
\end{figure}

Our program was designed to comprise two long ($\sim 100$\ ks each) 
uninterrupted observations of each field, separated by six months or 
more, although  in some cases, scheduling constraints dictated a larger 
number of shorter observations.   \chase\/ observations began in 
September 2005; in this paper we present the observations performed in 
the first year of observations (i.e., through August 2006).  
The observations are summarized in Table~\ref{obstab}.
At this point each field had been observed for roughly half of the 
200 ks total time, and two fields (4 and 6) have their
observations complete. In addition to the \chase\/ observations, we utilized 
two archival ACIS-I observations of \m\ (OBSIDs 1730 \& 2023), one centered on 
the nucleus and one centered on the bright \hii\/ region NGC~604.
 There are two other archival observations of \m\ which we
chose not to use since one (OBSID 786) was conducted with the ACIS-S 
array with its smaller FOV and the other (OBSID 787) was conducted in 
subarray mode with the ACIS-S array with an even smaller FOV.
 Some of the fields (1,2,3, second half of 4, and 7)
were executed in continuous $\sim100$~ks intervals.  However some of 
the fields 
were broken into several pieces with field 6 being the most extreme
example (6 observations, with one as short as 12~ks). 
To show the coverage, we display the exposure map of the \chase\/ observations 
combined with the archival ACIS-I observations 1730 and 2023 in 
Figure~\ref{expmap}. The combined exposure map is complicated, with the
minimum exposure being $\sim3.7\times10^{6}~{\rm s~cm^2}$ and the maximum 
exposure $\sim1.56\times10^{8}~{\rm s~cm^2}$. The pattern of overlapping 
exposures results in a non-uniform exposure map, with the lowest exposures
in the gaps between the CCDs and the highest exposure in the regions where
3 of the fields overlap.  \chandra\/ dithers while conducting observations, 
so that regions in the chip gaps receive some exposure (the lowest exposure
typically being half that of the immediately surrounding regions
unaffected by the dithered chip gap). 
Further complicating the exposure map is the fact that the
overlapping regions typically combine data from different off-axis angles
for which the response of the telescope is significantly different.
Both the effective area and the angular resolution of the telescope 
decrease with increasing off-axis angle making it undesirable to
simply add the data from overlapping observations. The implications of
these effects for source detection and characterization are 
discussed in \S~\ref{sec-catalog}.

\subsection{Data Processing}\label{sec-proc}

All \chase\/ observations are performed with ACIS-I as the primary instrument 
in `Very Faint' (VFAINT) mode. The data are reprocessed with CIAO 
version 3.2.2 and CALDB version 3.1.0. 
First, we filter on the background flags for the VFAINT mode to improve
the background rejection efficiency.
We then apply the charge transfer inefficiency (CTI) correction
to the archival observations (this correction was already applied to
the \chase\/ data in the standard processing) and  
the time-dependent gain correction appropriate for the
time interval during which the data were acquired.
The events file is then screened with grade and status filters, good time
intervals are selected, and pixel randomization is removed. 
Finally, we check for background flares by creating a background light
curve from the events on the ACIS-I3 chip,
after removing point sources.  
We perform an iterative sigma-clipping algorithm to remove time intervals with 
count rates more than $3\sigma$ from the mean of each iteration, until all 
count rates are less than 
$3\sigma$ above the mean.\footnote{http://cxc.harvard.edu/ciao/threads/filter\_ltcrv/}
Table~\ref{obstab} lists the effective exposure times after the light curves
have been filtered.
This filtering resulted in the selection of events which were used for the
analysis discussed in this paper.

\subsection{Mosaic Images}\label{sec-mosaics}

We created a mosaic image of the observations listed
in Table~\ref{obstab}, including the two archival pointings to increase
the exposure and provide additional coverage.  We selected events with 
off-axis angles $\le8.0\arcmin\/$ to include only the highest angular 
resolution data in these images and with energies $0.35 < {\rm E} < 8.0$~keV 
in order to maximize the X-ray signal compared to the background.
The data were binned into $2\arcsec\/\times2\arcsec\/$ pixels (a binning
of $4\times4$ on the original 0\farcs492 sky pixels).
The exposure maps are computed using the CIAO tools {\tt mkinstmap} and 
{\tt mkexpmap}.  For the instrument maps, we compute the spectral 
weights\footnote{ http://cxc.harvard.edu/ciao/threads/spectral\_weights/}
assuming a power-law spectrum with $\Gamma$ = 1.9 and 
\nh = $1.0\times10^{21}$~cm$^{-2}$.
We adopt an \nh\/ of  $1.0\times10^{21}$~cm$^{-2}$ based on the
estimates of the Galactic column density of $5.7\times10^{20}$~cm$^{-2}$
\citep{dickey1990} and $6.3\times10^{20}$~cm$^{-2}$ \citep{stark1992} 
and the total line-of-sight 
column density through \m\/  which averages about
$1.0\times10^{21}$~cm$^{-2}$ \citep{newton1980}.
We adopt a power-law model with $\Gamma$ = 1.9 as representative of
the intrinsic spectrum of AGNs \citep{pounds1994}.
The image is exposure-corrected and smoothed with a Gaussian filter 
with a $\sigma=3$ bins corresponding to 6\arcsec.  Figure~\ref{mosaic} shows 
this mosaic image for the energy band 0.35 -- 8.0~keV. 
The nucleus dominates the center of the image at 
\about\ RA(J2000) =
01$^\mathrm{h}$~33$^\mathrm{m}$~50\fs9~, Dec(J2000) =
+30\degr~39\arcmin\/~36\farcs8 (hereafter all coordinates are provided in 
epoch J2000).
There are numerous bright X-ray sources visible in this image, notable
among them is the XRB X-7 at 
RA = 01$^\mathrm{h}$~33$^\mathrm{m}$~$34\fs12$~, 
Dec = +30\degr~32\arcmin\/~11\farcs6\/. 
The diffuse emission is not as evident in this image as in the \xmmnewton\/
data \citep{pietsch04} due to the ACIS-I array's lower 
sensitivity to soft X-rays.
However, several bright diffuse objects are easily visible;
e.g., NGC\,604, the brightest \hii\/ region in \m\/,
northeast of the nucleus at \about\ 
RA = 01$^\mathrm{h}$~34$^\mathrm{m}$~34$^\mathrm{s}$~, 
Dec = +30\degr~47\arcmin\/, and  the \hii\/ region IC\,131,
northwest of the nucleus at 
\about\ RA = 01$^\mathrm{h}$~33$^\mathrm{m}$~15$^\mathrm{s}$~, 
Dec = +30\degr~45\arcmin\/.

We also created images in narrower energy bands (soft [0.35 -- 1.1~keV], 
medium [1.1 -- 2.6~keV], and hard [2.6 -- 8.0~keV]) to distinguish between 
softer and harder emission. The three-color mosaic image of all the 
ACIS-I observations of \m\ is presented in Figure~\ref{rgbmosaic}. 
The image shows different types of sources in different colors: e.g., most of 
the known SNRs appear red or orange, while hard sources like XRBs or
background AGN are blue or white. 
The nucleus appears white in the image since it is bright in all
three bands. 
The \hii\/ region NGC\,604 appears rather red while the \hii\/ region IC\,131
appears green, a clear indication that the spectrum of IC\,131
is significantly different from that of NGC\,604.
The three-color image also shows faint and soft diffuse emission 
northwest of the nucleus as well as extending from southeast to south  
which traces the southern spiral arm of \m\/.

\subsection{Source Catalog}\label{sec-catalog}

We searched for candidate sources using the CIAO tool {\tt wavdetect}.
Each \chase\/ observation or group of observations conducted at the
same pointing and roll angle was searched; we did not search the
archival observations.  For each ACIS-I chip, we binned to
0.492\arcsec{} sky pixels and ran {\tt wavdetect} for the band 0.35--8 keV.
We used a significance threshold of $1.0\times10^{-6}$ which corresponds
to a false detection probability of about 1 source per CCD
\citep{freeman02}.
To reduce spurious source detections at the chip edges, we used
an exposure map (evaluated at 1.5 keV) and required that the exposure
be at least 10\% of the peak value.  In order to be sensitive to spatial
scales up to \about~0.5\arcmin, we used a power-of-2 sequence of scales
from 1 to 64 pixels.  Only detections with a significance of 
$> 3\sigma$\ as determined by {\tt wavdetect} were considered for
further investigation.  The signal-to-noise ratio computed by
{\tt wavdetect} is the net counts divided by the \citet{gehrels86}
estimate for background error $\sigma_B = 1 + \sqrt{C_B + 0.5}$,
where $C_B$ is the estimate of background counts in the source
detection region.
These detections were used as input source positions for the software
package {\tt ACIS Extract} \citep{Broos02}.

{\tt ACIS Extract} is a multi-purpose source characterization package
which computes source and background count rates and fluxes in a group of
user-specified energy bands, computes source significance, extracts
spectra and generates appropriate response files, among other
items \citep[see][for details]{Broos02}.
Of particular interest and advantage for our analysis is the fact that
{\tt ACIS Extract} produces an extraction region tailored to match
the \chandra\/ point spread  function (PSF) at the position of the
source in each observation. This
capability is crucial for our analysis since we have multiple,
overlapping FOVs in which a particular source may end up at a variety
of off-axis and azimuthal positions.  Each of these positions requires
a different extraction region based on the PSF to optimize the source
signal compared to the background, since the \chandra\/ PSF is a
function of both off-axis angle and azimuthal angle. The software
produces a background region for each source that excludes the source
extraction regions of nearby sources.
{\tt ACIS Extract} allows a visual inspection of each source in each
exposure along  with an outline of the PSF. These visual inspections
allowed us to flag clearly extended
sources in the catalog and to remove duplicate sources that were observed in
more than one observation in overlapping sections of fields. We also
identified observations in
which the source extraction region was affected by the ``transfer
streak'' of the readout of the ACIS CCDs.  Typically a source was
affected by the transfer streak in only one observation (or group of
shorter observations conducted at the same pointing position and roll angle).
We then excluded those observations from the source characterization
while retaining the observations in which the source was unaffected by
the transfer streak. The inspections also allowed us to adjust the
source positions in cases where there were multiple detections and the
{\tt wavdetect} source position chosen did not match the centroid of
the most nearly on-axis observation of the source.

The final judgment on
the significance of a source was based upon the source and background
counts calculated by {\tt ACIS Extract} in the 0.35--8.0 keV band.
The signal-to-noise ratio calculated by {\tt ACIS Extract} uses the
difference between the 
\citet{gehrels86} estimate for the Poisson {\it upper} limits on 
the source and background counts and the net extracted counts; this 
generally results in a lower significance than  {\tt wavdetect}.  In
this manner, we used {\tt wavdetect} as a candidate source {\it detection}
tool and used {\tt ACIS Extract} as a source {\it characterization} tool.
Our approach is conservative in two respects. First, for the two
fields in which the first and second epoch observations were 
available (fields~4 and~6), we did not add the data and conduct a 
source detection on the summed data. We plan to conduct such an
analysis on the full survey data and we expect that fainter sources
will be detected in such an analysis.  Second, we filtered twice on
the source significance, once on the output of {\tt wavdetect} and a
second time on the output of {\tt ACIS Extract}. It is conceivable
that there are sources which were less significant than $3\sigma$ in
the {\tt wavdetect} output which would have become more significant than 
$3\sigma$  in the {\tt ACIS Extract} output after the source and
background extraction regions had been adjusted.  We decided to be
conservative in this first catalog 
and will explore alternative source detection techniques
in a future paper.

We estimated the extended-source positions using an iterative
``$\sigma$-clipping'' algorithm as described in \citet{Ghavamian05}.
Based on an initial position estimate and clipping radius, the centroid
of the events (0.35--2.6\,keV) within the clipping radius is 
evaluated.  Events more than $n\sigma$ ($n=1.5-2$) from the 
centroid are removed, and a new centroid
is evaluated.  The procedure is repeated until convergence within
0.01 pixels is attained, or for a maximum of 10 iterations.
For each extended source, this procedure was applied for selected
observations in which the source was closest to on-axis, or contained
the largest number of counts.  We examined the exposure maps, and
only included those observations for which the exposure was fairly
uniform across the centroiding region (avoiding observations with
the source near a chip edge).  In many cases, more than one observation
qualified, in which case we estimate a weighted mean position, where
we weight by the net exposure time at that location on the CCD
for the observation.

Finally, the {\tt ACIS Extract} value of the source significance was
applied to remove source candidates with less than $3\sigma$ detections in the
0.35-8.0 keV band, therefore forcing all of the objects in our final
catalog to show $3\sigma$ significance by both measuring techniques.
We measured all of the fluxes, spectra, and timing for our final
source list using {\tt ACIS Extract}.  The {\tt ACIS Extract}
measurements include short-term and long-term variability information as
well as fluxes.  A variability study of the sources is in progress 
and will be addressed in future publications 
(Williams~\etal\/~2007, private communication).

Table~\ref{newtab} is the source
catalog of the \chase\/ data containing 394 sources. 
The columns in Table \ref{newtab} contain
the source number, the position in RA(J2000) and Dec(J2000), 
the error in the position, the net counts after background subtraction,
the count rate in
the 0.35--8.0~keV band with the error in the count rate listed in
parentheses, the photon flux (absorbed) in the
0.35--8.0~keV band with the error in the flux listed in parentheses,
and the hardness ratios HR1 and HR2 as defined in \S~\ref{sec-HRs}
with the appropriate errors listed in parentheses after each HR
value. 
Extended sources are indicated by a superscript {\it e} on the 
source number; 23 sources show evidence of spatial extent beyond the 
local PSF.
Given the source selection criteria described above, we estimate that
the faintest source we could detect in a 100~ks observation, close to
on-axis would have $\sim15$ net counts.  Source \#385 is the source
with the lowest number of net counts in our list with 16.4 net counts.
The minimum number of counts necessary for a $3\sigma$ detection
grows with off-axis angle as the PSF increases.  We estimate that a
source 8\arcmin\/ off-axis would need $\sim30$ net counts to satisfy
our source detection criteria. 
The photon fluxes were calculated in the three bands 0.35-1.1 keV,
1.1-2.6~keV, and 2.6-8.0~keV, using the net counts in the band, a
mean effective area for that band, and the exposure time. The mean
effective area for a band was computed assuming a flat spectrum,
this partially accounts for the fact that the \chandra\/ effective
area is a strong function of energy. The photon fluxes in the
individual bands were then summed to produce the photon flux
in the 0.35-8.0 keV band.  Therefore, we have not made any assumptions
about the spectra of the sources to convert the detected counts into
a photon flux, other than the assumption of a flat spectrum for the
mean effective area. For illustrative purposes, we will now assume a
model spectrum for the lowest and highest flux sources in our list 
and perform spectral fits in {\tt XSPEC} to
quantify what the range of energy fluxes might be for these assumed
spectral models.
The lowest flux source in the list is source \#315, with a
photon flux (0.35--8.0~keV) of $2.50\times
10^{-7}$ photons cm$^{-2}$ s$^{-1}$ which
corresponds to an energy flux (0.35--8.0~keV, absorbed) of
$8.4\times 10^{-16}$ erg cm$^{-2}$ s$^{-1}$ assuming a power-law
spectrum with a photon index of 1.9 and an \nh\/ of 
$1.0\times10^{21}~{\rm cm^{-2}}$.  This flux implies a
luminosity (absorbed) of $6.7\times 10^{34}$ erg s$^{-1}$ at
the distance of \m.  Even with the low number of counts, the reduced
$\chi^2$ of the fit improves significantly when the \nh\/ is allowed to
vary, dropping from 1.51 to 0.48 for an \nh\/ of 
$9.1\times10^{21}~{\rm cm^{-2}}$.  The fit with this model implies
a luminosity (absorbed) of $1.6\times 10^{35}$ erg s$^{-1}$,
$2.4\times$ higher than the estimate with an \nh\/ of 
$1.0\times10^{21}~{\rm cm^{-2}}$.
The highest flux source is of course the nucleus
(\#200 in this list) with a photon flux of
(0.35--8.0~keV) of $1.95\times 10^{-3}$ photons cm$^{-2}$
s$^{-1}$.  A spectral fit with an absorbed 
power-law model returns an $\nh=4.5\times10^{21}~{\rm cm^{-2}}$,
an index of 1.99, an energy flux (0.35--8.0~keV, absorbed) of $5.1\times
10^{-12}$ erg cm$^{-2}$ s$^{-1}$, and a luminosity (absorbed) of
$4.1\times 10^{38}$ erg s$^{-1}$.
We note that this source is significantly
piled-up and the measured flux and luminosity are underestimates of
the true values.  The analysis of the nucleus and other sources
significantly affected by pileup will be discussed in a future paper.

\section{Analysis and Discussion}\label{sec-analysis}

The primary product of our analysis has been the creation of the source
catalog shown in Table~\ref{newtab}.  We now make use of that
catalog to begin to characterize the sources through their hardness
ratios, and to identify some of the sources by
cross-correlation with other catalogs.  In addition to
the results based on the source catalog, we present two examples of
the types of analyses which this rich data set allows,  namely
preliminary results on the \hii\/ region NGC\,604 and the southern arm
region.

\subsection{Hardness Ratios}\label{sec-HRs}

\begin{figure*}
\centering
\plotone{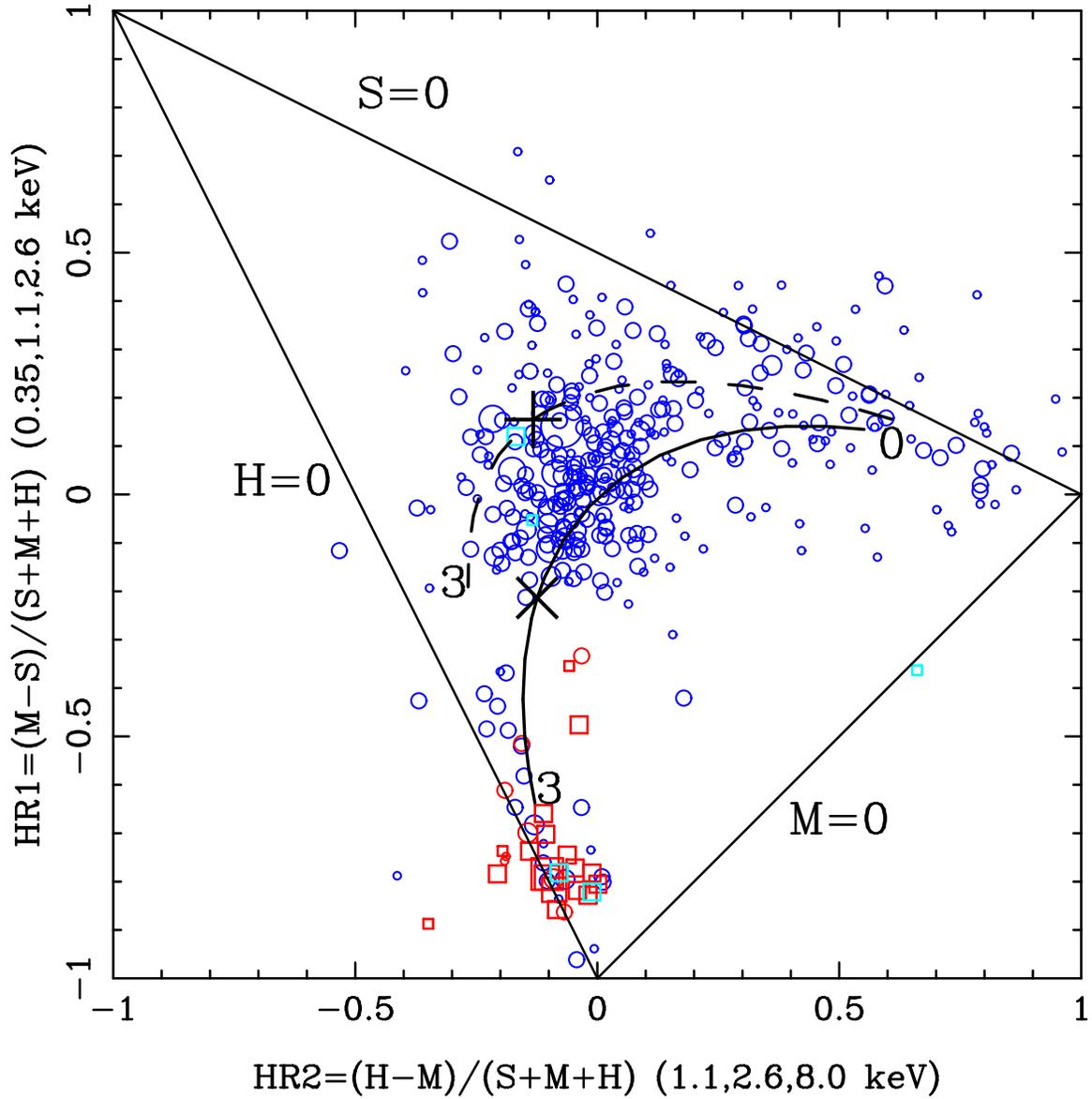}
\caption{\label{colorcolor}
Hardness radio plot constructed from the source list in Table \ref{newtab}.
Unidentified sources are plotted in blue and GKL98 sources are plotted
in red. Extended sources which are not in the GKL98 list are plotted
as cyan. Sources are plotted as circles except for extended sources,
which are plotted as squares.  The symbol sizes indicate the
approximate photon fluxes 
($\mathrm{photons}\,\mathrm{cm}^{-2}\,\mathrm{s}^{-1}$):
$< 10^{-6}$ (smallest symbols), 
$10^{-6}$ to $10^{-5}$,
$10^{-5}$ to $10^{-4}$, 
$10^{-4}$ to $10^{-3}$, and
$>10^{-3}$ (largest symbols). The HR values are based on the photon
fluxes and assume a flat spectrum to compute the mean effective area in the 
three bands.   The black curves indicate the loci of
power-laws with $\Gamma=0$--3 and with different $N_\mathrm{H}$:
solid for $N_\mathrm{H} = 1\times 10^{21}\,\mathrm{cm}^{-2}$,
and dashed for $N_\mathrm{H} = 3\times 10^{21}\,\mathrm{cm}^{-2}$.
The large black $\times$ (+) indicate the $\Gamma=1.9$ powerlaw cases.
}
\end{figure*}

Using the background-subtracted photon fluxes for the \chase\/ sources,
we created a hardness ratio diagram 
(Fig.~\ref{colorcolor}).  
We construct hardness ratios using a definition similar to that
described in \cite{prestwich2003} from background-subtracted fluxes for
soft ($S$), medium ($M$), and hard ($H$) bands:
\begin{equation}
\mathrm{HR1} = {(M-S) \over (S+M+H)}, 
\end{equation}
and
\begin{equation}
\mathrm{HR2} = {(H-M) \over (S+M+H)}, 
\end{equation}
where in our case, $S = 0.35$--1.1\,keV, $M = 1.1$--2.6\,keV,
and $H = 2.6$--8\,keV (the same bands used for the three-color image in
\S\ref{sec-mosaics}). We estimate the uncertainties on the fluxes
  in a given band by taking the differences between the 
  \citet{gehrels86} upper limits and the net counts for the band.
  The uncertainties for the hardness ratios are then estimated 
  by propagating the flux errors assuming Gaussian statistics (the
  uncertainties are included in parentheses behind the HR value in 
  Table~\ref{newtab}). 

The hardness ratio HR2 is plotted against HR1 in Fig.~\ref{colorcolor}.
    We indicate the limiting fluxes by solid black lines: $S = 0$
    (upper left to right center), $M = 0$ (right center to bottom center),
    and $H = 0$ (bottom center to upper left).  We plot all the sources,
    including those listed with zero or negative fluxes in some bands.
    Most of the latter lie near the boundary of the allowed region, and are
    clearly real sources with spectra similar to those just within the
    allowed region. If the true flux in a band were zero, the uncertainties
    associated with background subtraction would yield a negative observed
    flux roughly half the time.  We note that \cite{hong2004} and 
    \cite{park2006} have
    suggested that the ``quantile'' method and Bayesian approaches  
    have significant advantages
    over the traditional hardness ratios method for faint sources 
    which have a low number of counts in a given band. 
    We will consider the use of these new methods in our final 
    catalog paper.
    The source hardness ratios are plotted
    as symbols with sizes indicating the 0.35--8\,keV photon fluxes
    (photons\,cm$^{-2}$\,s$^{-1}$): $\le 10^{-6}$ (smallest symbols),
    $10^{-6}$--$10^{-5}$, $10^{-5}$--$10^{-4}$,
    $10^{-4}$--$10^{-3}$, and $> 10^{-3}$ (largest symbols).
    GKL98 objects (discussed below) are plotted with red symbols,
    extended objects which are not in the GKL98 list are plotted as
    cyan, and 
    the rest are plotted in blue.  In each case, apparently nonextended 
    sources are plotted as circles, and extended sources are plotted as 
    squares.  For reference, we plot loci for absorbed
    powerlaws computed using {\tt XSPEC} (again the HR values are based on
    the photon fluxes in \S~\ref{sec-catalog} which assume a flat
    spectrum to compute the mean effective area in the three bands).  
    The black solid 
    curve shows the locus for powerlaws with absorption
    $N_\mathit{H} = 10^{21}\,\mathrm{cm}^{-2}$, and powerlaw index
    $\Gamma$ ranging from 0 to 3; the large black ``$\times$''
    symbol indicates the $\Gamma = 1.9$ powerlaw.  The black dashed
    curve is similar except that a larger absorbing column,
    $N_\mathit{H} = 3\times 10^{21}\,\mathrm{cm}^{-2}$, is assumed.
    The large black ``$+$'' symbol indicates the $\Gamma = 1.9$ powerlaw
    for this case.

    We expect the large clump of sources in the center of the diagram 
    will prove to be a combination of XRBs in M33 and background AGN.
    The nucleus is indicated by the large blue circle near the center 
    of the diagram. 
    We also detected objects associated with the GKL98 list of
    optically-identified SNRs, and some
    clearly extended sources.  The sources identified with GKL98 objects
    are plotted with red symbols, again using circles for nonextended 
    sources (18 objects), and squares for extended sources (8 objects).
    Finally, in addition to the GKL98 extended objects, we detect an
    additional 5 objects which do not appear in the GKL98 list; these are
    plotted as blue squares.  As expected, the GKL98 supernova remnants
    tend to cluster in the soft portion of the diagram (bottom center,
    near the $H=0$ line).

\begin{figure*}
\centering
\plotone{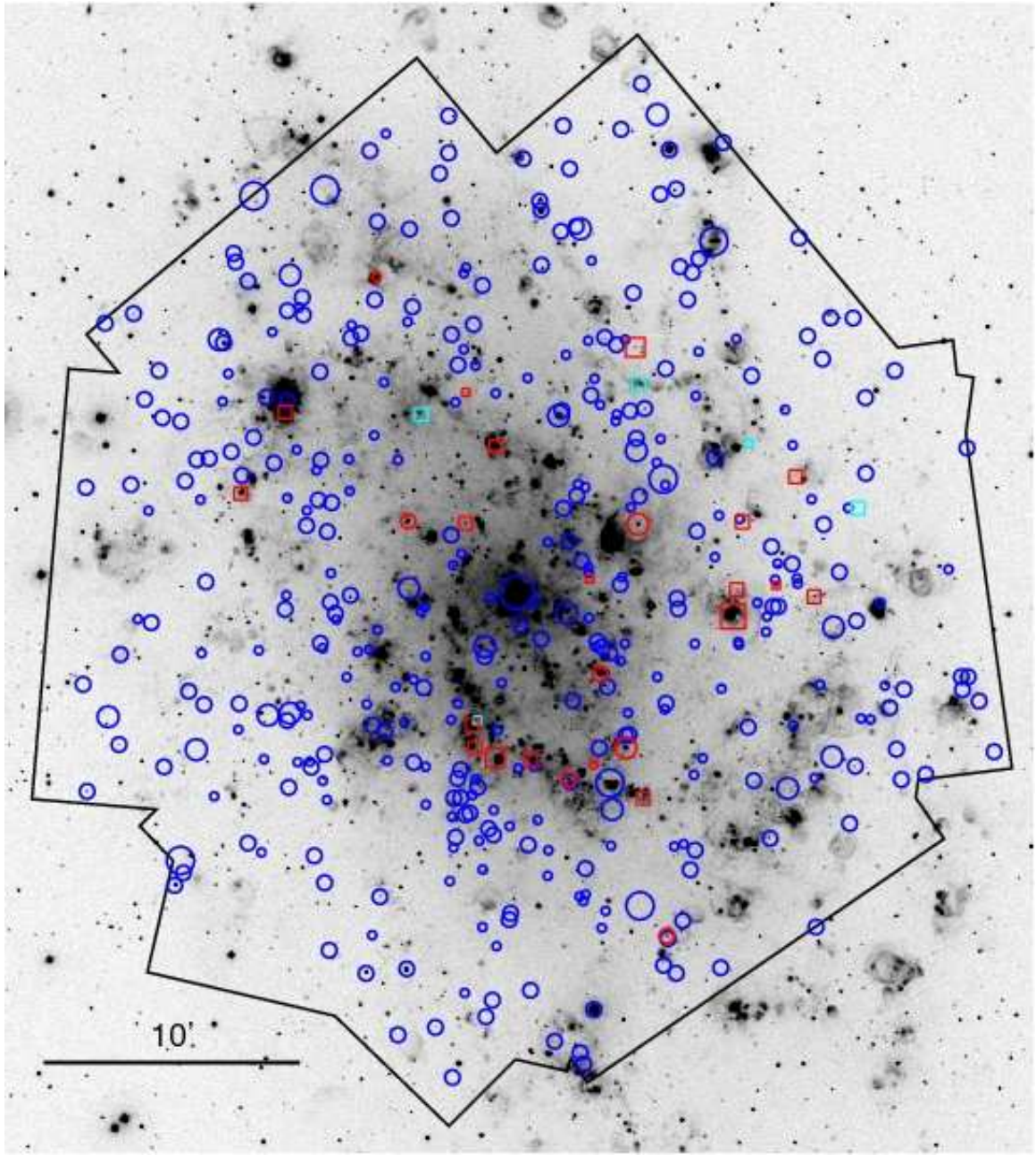}
\caption{\label{srcsonhalpha}
Survey sources plotted on a deep \halpha{} image taken with the
Burrell Schmidt telescope at Kitt Peak (McNeil \& Winkler 2006).
Unidentified sources are plotted in blue, GKL98 sources are plotted
in red and extended sources not in the GKL98 list are plotted in
cyan; extended sources are plotted as squares, and the
rest are plotted as circles.  The symbol sizes encode the photon fluxes
(see Fig. 5).  The black polygon outlines the overall FOV of the
overlapping ACIS-I FOVs of the survey.
}
\end{figure*}

   In order to explore the distribution of the \chase\/ sources across
the galaxy we have plotted the positions of the sources on a 
\halpha\/ image \citep{2006AAS...39....80M} in
Figure~\ref{srcsonhalpha}.  We have used the same symbols and colors
as those used  in the hardness ratio diagram in
Figure~\ref{colorcolor}.  The concentration of SNRs along the southern
spiral arm is evident in this image. Also interesting is the clump
of unidentified sources (indicated by the medium-size blue circles)
just south of the concentration of SNRs. This region warrants a closer
examination in the future.  The nucleus is again clearly indicated by
the large blue circle in the center of the image.  We note that
at least half of the sources in the \chase\/ catalog and this
image are expected to be foreground and background sources.  One would
expect the
foreground and background sources to be uniformly distributed across
the region of the survey.  Therefore, any concentration of sources
aligned with structure in the galaxy, such as the southern spiral arm,
is more likely to have a higher percentage of \m\/ sources.
Identifying which sources are related to \m\/ and those which are
unrelated is a non-trivial task which will require a significant
effort to conduct optical followups of these sources. A careful study
of the characteristics of the X-ray source population in \m\/ will
require such an identification and an accurate means of estimating
the contribution of foreground and background sources. 


\def\OIGS{\:{\rm ergs\:cm^{-2}\:s^{-1}\:\AA^{-1}}}
\newcommand{\MSOL}{\mbox{$\:M_{\sun}$}}  
\newcommand{\EXPN}[2]{\mbox{$#1\times 10^{#2}$}}
\newcommand{\EXPU}[3]{\mbox{\rm $#1 \times 10^{#2} \rm\:#3$}}  
\newcommand{\POW}[2]{\mbox{$\rm10^{#1}\rm\:#2$}}


\subsection{Comparison with Previous Catalogs}\label{sec-IDs}

In order to gain some perspective on the sources contained in this 
catalog of X-ray sources in M33, we have cross-correlated the \chase\/ 
sources with sources detected in previous X-ray studies  as well as 
with a number of catalogs constructed from optical and radio studies of \m\/. 
A total of 17 sources were identified in the vicinity of \m\/ using 
{\it Einstein}  \citep{trinchieri88}.  Of these, 13, including all 
of the brighter sources, lie within 2\arcsec\/ of \chase\/ sources.  
The sources 
that were missed by us were either outside of the region that was 
surveyed in \chase\/, were detected only with the Imaging Proportional 
Counter, or were detected at very low statistical significance in the
{\it Einstein} data.  Excluding for the moment other \chandra\/ observations, 
the deepest X-ray surveys of \m\/ are those 
carried out by PMH04 and 
\citet[][hereafter MPH06]{misanovic06} using \xmmnewton\/.  
MPH06 took the most care in using data from individual observations
from which one 
could obtain accurate source positions and their catalog contains 350 
sources with typical position errors of 2\arcsec\/ for the brighter
sources. PMH04  
included more data and their catalog has 408 sources, 97 of which are not 
contained in the source list compiled by MPH06.  Of the 
sources in the PMH04 and MPH06 catalogs, 225 and 189 
of the sources are found in regions covered by the \chase\/ survey  
(at an effective exposure exceeding \POW{6}{s~cm^{-2}} at 1 keV). 
The 3$\sigma$ position
errors for the faintest sources in both \xmmnewton\/ surveys are about 
7\arcsec\/.  Assuming this positional 
uncertainty, there are 198 and 154 \chase\/ sources that have counterparts 
in the two surveys, indicating a high degree of overlap.
The source correlations are listed in
Table~\ref{source_ids}.   Not surprisingly, 
given the higher resolution of \chandra\/, there are cases where multiple 
\chase\/ sources are identified with a single \xmmnewton\/ source.    Of the \xmmnewton\/ 
sources that are in the region we surveyed , there are 28 and 26 sources, 
respectively, that were not detected.
A superficial inspection of the \xmmnewton\/ 
sources that were not detected shows that they tend to be the fainter sources, 
with a median count rate 2.5 times lower than the \xmmnewton\/ sources that have 
counterparts.  They also tend to have softer X-ray spectra than the 
sources that have counterparts; for example, in the PMH04 list, 
the sources without counterparts have a mean (0.5-1 keV  vs 1-2 keV) 
hardness ratio  of  -0.14, compared to +0.15 for the sources with 
counterparts, which may reflect \xmmnewton\/'s greater effective area
at low energy compared with ACIS-I.

\citet[][hereafter GMZ05]{Grimm05} created a source catalog
from their analysis of 
3 earlier \chandra\/ observations of M33, two centered on the nucleus of M33, 
and one centered on the star-forming \hii\/ region, NGC604.  They report a 
total of 261 sources; based on a detection using {\tt wavdetect} in any of the 
3 observations and in any of 3 bands (soft: 0.3--2.0~keV, hard:
2.0--8.0~keV, and total: 0.3--8.0~keV), with positional uncertainties that 
are typically $<2$\arcsec\/, but are occasionally considerably larger
(up to 10\arcsec\/).  
Of these 261 sources, only 145 have counterparts in the \chase\/ source 
list allowing a position uncertainty of 3\arcsec\/, and 99 remain unidentified 
assuming a position uncertainty of 10\arcsec\/.  This difference 
clearly demands 
explanation, especially since the \chase\/ exposures are deeper than (and 
in some cases overlap with) those analyzed by GMZ05.  Of the 261 
sources identified by GMZ05, 249 are in regions where the effective 
exposure exceeds \POW{6}{s~cm^{2}} at 1 keV, and so disjoint fields of view 
do not explain the difference. Instead, the  principal reason for this 
difference is that GMZ05 used far less conservative source selection 
criteria than those we have used, detailed in section \ref{sec-catalog}.
While our source selection required detections of at least
$3\sigma$ significance in both {\tt wavdetect} and {\tt ACIS
Extract}, their selection criteria demanded only a detection
by {\tt wavdetect} at any significance. 
Of the 261 sources listed by GMZ05 there are 
144 which have fluxes which are measured at $3\sigma$ in any one of the 
observations analyzed by GMZ05; of these 112 are matched assuming 
a position uncertainty of 3\arcsec\/, rising to 122 if a position uncertainty 
of 10\arcsec\/ is allowed.  Some of the remaining differences are likely due 
to the time 
variability of the sources.  If we restrict ourselves to the 29 sources 
with count rates determined at $10\sigma$ detected by GMZ05, there 
are still 4 which are unmatched assuming a position uncertainty of
10\arcsec\/; of these, two 
(J013435.1+305646 and J013444.6+305535 ) were not in regions included in 
the \chase\/ data, one (J013433.7+304701) is located in NGC604 and we 
excluded it as a separate source, and the remaining source appears to be 
time-variable.

An important goal of the \chase\/ survey is to study 
the X-ray properties of the SNR population in M33.   The most extensive 
catalog of SNRs in M33 remains the catalog of GKL98, which 
contains 98 SNRs identified on the basis of strong 
[\ion{S}{2}]:\halpha\/ emission.  \cite{Ghavamian05} identified two more
SNRs based on elevated [\ion{S}{2}]:\halpha\/ ratios and the soft
X-ray spectra from \xmmnewton\/, bringing the total to 100.
After updating the positions of a few of the SNRs based on our interference
filter 
imagery, we find  that 28 lie within 10\arcsec\/ of \chase\/ source 
positions.  
Nearly all of these sources have very soft X-ray spectra, consistent with 
their identification as SNRs.  
With two exceptions, a comparison of 
X-ray surface brightness distribution and \halpha\/ images in the 
vicinity of each of these sources also supports their identification as 
SNRs. The exceptions are \chase\/ source \#222, which is close to GKL98-59, 
and \chase\/ source \#242, which is close to GKL-66. Both of these sources 
appear point-like, and lie outside the optical nebulosity associated 
with the SNR. Source \#242 is also the only candidate SNR that has a hard 
X-ray spectrum.
 We \citep{Ghavamian05} had previously 
searched M33 for SNRs using \chandra\/ archival data; in that search, 21 SNRs 
were found to match objects in the GKL98 list.  Of these, 16 are 
contained in our list of SNRs with X-ray counterparts.  However, there 
were 5 SNRs in our earlier list that were not confirmed here.  This is 
almost surely due to the fact that, in the earlier search, we used band-passes 
that were optimized for SNRs rather than the broad-band used here. 
 Thus, excluding the two doubtful associations discussed above, 
 the total number of optically identified SNRs detected with 
\chandra\/ now totals 31, roughly 1/3 of all of the SNRs that are known in 
\m\/. PMH04 performed an initial search of the \xmmnewton\/ data 
for SNRs using criteria similar to our own.  They identified 21 of the 
GKL98 SNRs.  In their reanalysis of the data, MPH06 confirmed 15 of
these, and suggested that 3 others 
(corresponding to SNRs 21, 28, and 55) were not SNRs since they exhibited 
time variability.   All of the latter are currently included in our list 
of identifications, though it is clear that further studies are warranted.  
In the \chandra\/ data, SNR21 is resolved as an elliptical shell with a 
size of 5\arcsec\/; 
it was the subject of a detailed discussion \citep{gaetz07}.

The USNO3 catalog contains about 8600 
``stars'' in the region that was surveyed as part of \chase\/, 69 of which 
lie within 3\arcsec\/ of \chase\/ sources.  The majority of these are 
not single 
stars, but instead represent a heterogeneous collection of objects, 
including stellar associations, \hii\/ regions, and background galaxies.  
Given this number of objects and an allowed error of 3\arcsec\/, one expects 
about 24 chance coincidences.  Therefore, although many of the 
correspondences likely represent a physical association, the 
effort to determine which are real and which are not is beyond the scope 
of this initial discussion of the \chase\/ sources.  Therefore in 
Table~\ref{source_ids}, 
we have identified only the 12 objects with an {\it R} magnitude of
less than 15 that lie within 3\arcsec\/ of a \chase\/ X-ray source;
there are 436 USNO3 sources with an {\it R} magnitude of 
less than 15 and thus only one of these associations 
is expected to occur 
by chance.  A number of these sources have already been identified as stars
 by PMH04 or MPH06, and 
\citet{Hatzidimitriou06} had already studied the counterparts of 6 of 
these sources. 

Using the {\it Ultraviolet Imaging Telescope}, 
\citet{massey96} identified 356 UV bright sources brighter than 
\EXPU{2.5}{-15}{\OIGS}, located preferentially along the spiral arms, 
and nearly all of which (351) are in the region surveyed in \chase\/.  
The objects in this list contain many O and early B-type supergiants, as well 
as a number of "super luminous"  Wolf-Rayet stars and OB associations.   
Ten of these sources are located within 5\arcsec\/ of 
\chase\/ X-ray sources, 
of which only one is expected by chance, and these are listed in 
Table~\ref{source_ids}.  
This is an indication that the X-ray source population in M33 is associated 
strongly with star formation. We also compared the \chase\/ source list to 
a variety of other catalogs, including the lists of  star cluster candidates 
in M33 compiled by \citet{sarajedina07},  of planetary nebulae in 
M33 by \citet{ciardullo04}, of emission line objects by 
\citet{calzetti95}, and of giant molecular clouds by
\citet{engargiola03}.  
Although there were a small number of coincidences, in nearly all 
cases the number of coincidences was close to that expected by chance.  
This does not imply that there are no physical associations with these 
objects in specific cases, but does imply that much more work will be 
required to isolate the subset of sources that have a physical association 
with these source populations.  A more detailed analysis of such 
correlations will be considered when the full \chase\/ dataset has
 been analyzed.

\begin{figure*}
\centering
\plotone{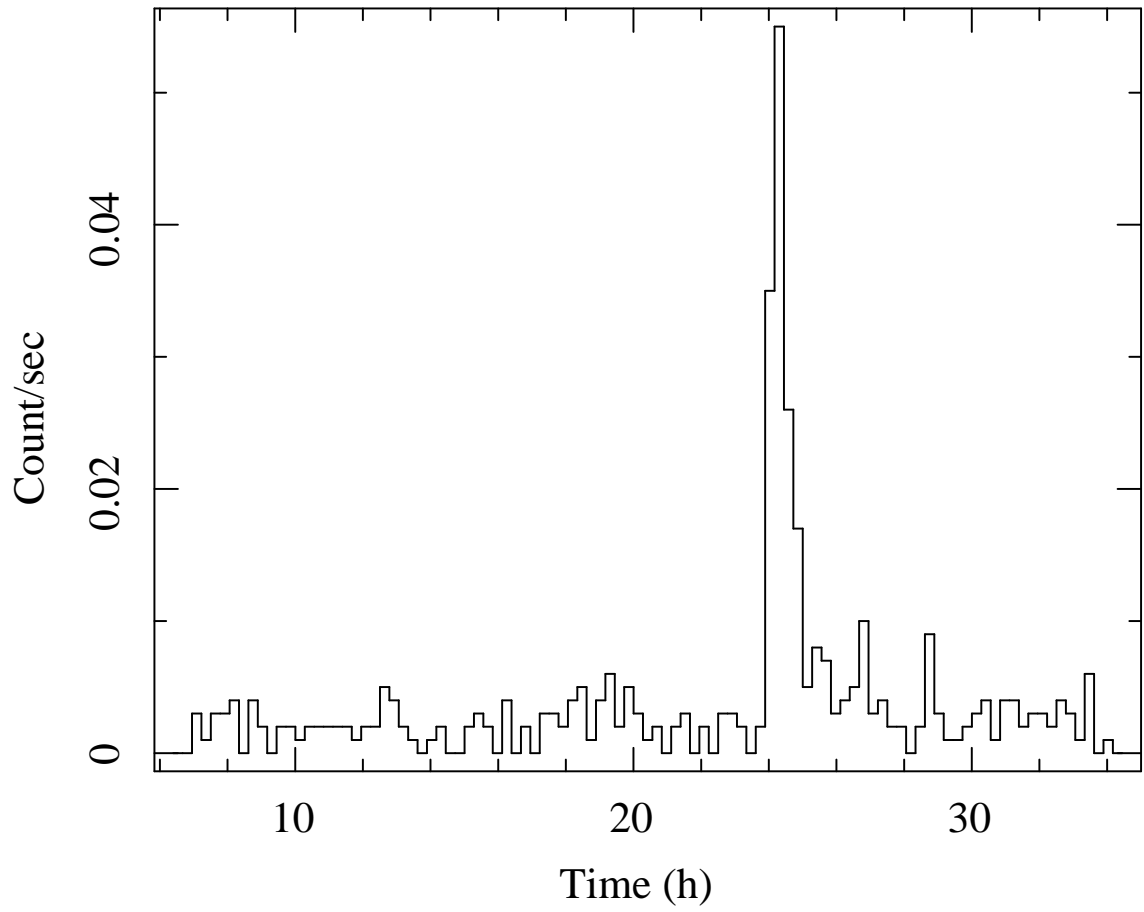}
\caption{\label{lc_star}
\chandra\/ ACIS-I light curve of \chase\/ \#175 ([PMH04] 196)
in the 0.5--2.5~keV band during ObsID 6383. Time zero corresponds to 
15 June 2006 .
}
\end{figure*}

Besides position and hardness ratio information, time variability arguments
can be used to strengthen identifications. An interesting source in
this respect is \chase\/ \#175 ([PMH04] 196) only ~1\arcmin\/~west
southwest of the 
bright nuclear source X-8. Optical colors, as
well as X-ray hardness ratio, and ${\rm f_\mathit{x}/~f_\mathit{opt}}$, 
suggested an identification with an M-type star in the foreground. 
This identification has been
strengthened by X-ray flux variability detected in \xmmnewton\/ observations
\citep{misanovic06}. The optical spectrum allows the identification of
the optical counterpart as a late M star with \halpha\/ emission 
indicating a flare
star, and indeed  X-ray flaring was detected in early \chandra\/ ACIS
observations of the source \citep{Hatzidimitriou06}.
\chase\/ \#175  was in the field of view of several ChASeM33 pointings which
allowed us to search for flares in a much larger dataset.  While in many
observations little or no variability is present, a strong flare is detected in
ObsID 6383 on 16 June 2006  with a FWHM of ~2000s
(Fig.~\ref{lc_star}). During the
flare the source reaches a count rate of more than $7\times10^{-2}$
cts s$^{-1}$ in the 0.5--2.5 keV band. In total ~400 cts (~150 during the
flare) were collected from the source.

\subsection{Resolving Confused Regions}

Several regions in \m\/ have extended X-ray emission confused 
with point sources and/or high  source density.  Attaining a better 
understanding of these regions was a prime motivation behind the
\chase\/ observations.  Here we present our first 
analyses of two such regions: NGC~604 and the southern spiral 
arm.

\subsubsection{The Giant \hii\/ region NGC\,604}\label{sec-NGC604}\label{sec-ngc604}

\begin{figure*}
\centering
\plotone{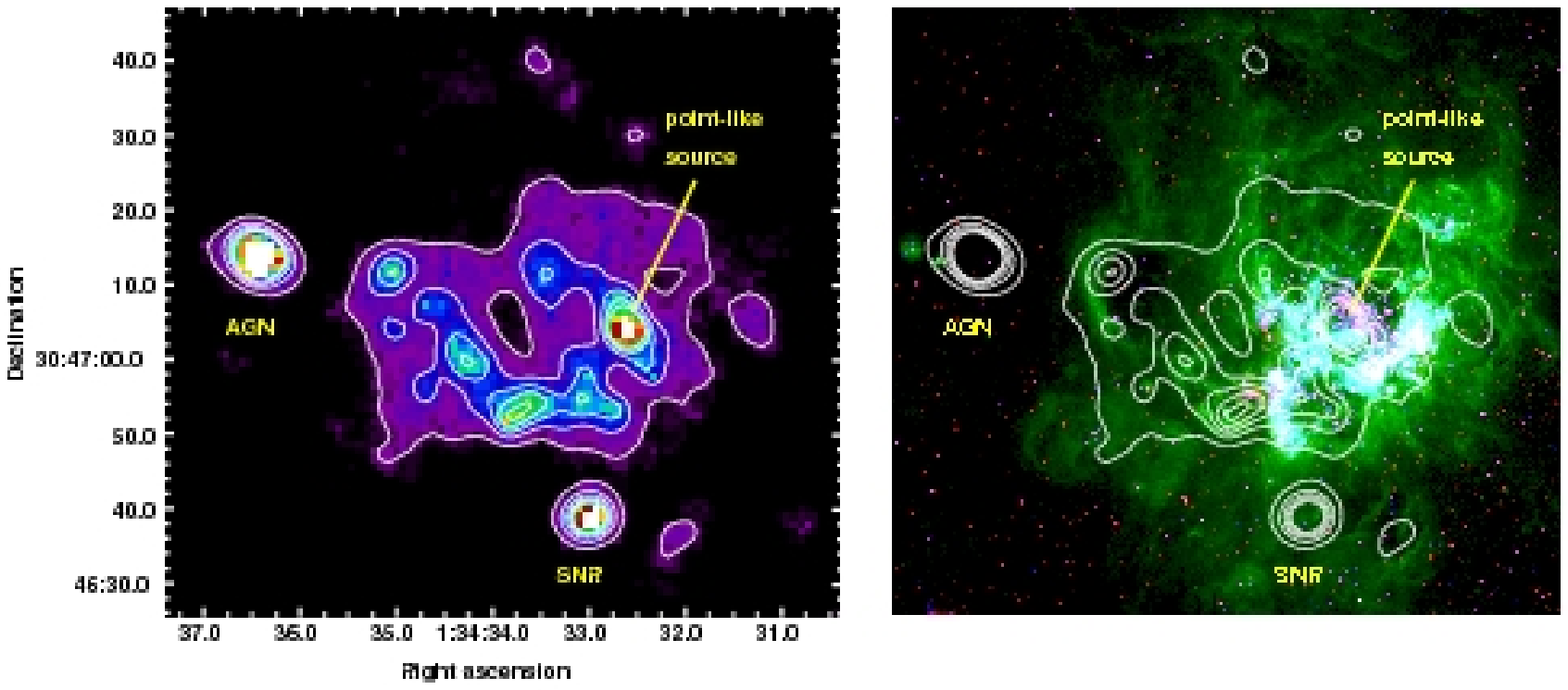}
\caption{\label{ngc604}
\chandra\/ ACIS-I image (left) and \hst\ WFPC2 images (right) of the 
giant \hii\/ region NGC\,604 with \chandra\/ contours 
(0.7, 1.2, 1.7, 2.2, and 2.7 counts/bin).  The images cover 
$1\farcm74\times1\farcm35$ region and 10\arcsec\/ corresponds to
$\sim40$~pc at the distance of \m.
The \chandra\/ image is binned with a bin size of 2 pixels and smoothed using 
the {\tt FTOOL fadapt} with a minimum of 20~cts in the kernel. In the
right panel, \hst\ $I$ band image is 
displayed in red, the \halpha\/ image in green, and the $V$ band image in blue.
}
\end{figure*}

An example of the power of the high-resolution images of \m\/ from
\chase\/  is the complex X-ray structure within and around 
the giant \hii\/ region NGC\,604, as shown in Figure \ref{ngc604}.
The left image shows 
the \chandra\/ ACIS-I image, binned into $2\arcsec\times2\arcsec\/$ pixels, 
with the X-ray contours overlaid on the data. 
In the right panel we show a three-color image from archival \hst\ WFPC2
observations {\citep[see][]{maiz04}}, with the
$I$ band in red (F814W, 200~s), \halpha\/ in green (F656N, 1000~s), 
and $V$ band in blue (F555W, 200~s),  with the \chandra\/ X-ray contours 
 overlaid on the data. 
The \chandra\/ image shows the hot diffuse gas inside the \hii\/ region. 
There is bright X-ray emission from a point-like source inside NGC\,604 
(in projection), southeast of the bright cluster of stars. The \chandra\/ 
image has also resolved a SNR (GKL98-94 and \chase\/ source \#~342)
located south of the \hii\/ region. The source to the east is probably a 
background source, as discussed below.  This image 
demonstrates how essential the high angular resolution of \chandra\/
is in separating the X-ray emission of the three discrete
sources from the diffuse emission of NGC\,604.

We perform spectral analysis of the prominent X-ray sources in this region 
using the \chase\/ data.
The diffuse emission in NGC\,604 is well fit with a thermal emission
model ({\tt APEC} in {\tt XSPEC}) with
an absorbing foreground column density of 
\nh\ = (0.53$\pm$0.02) $\times 10^{21}$~cm$^{-2}$, 
a temperature corresponding to $kT = 0.195\pm0.007$~keV,
and a flux (absorbed) of $F_{\mathrm X}$(0.5-2~keV) = $5.0 \times 
10^{-14}$~ergs~cm$^{-2}$~s$^{-1}$; 
this does not include the bright point-like source inside NGC\,604.
The point-like source is located at 
RA = 01$^\mathrm{h}$~34$^\mathrm{m}$~$32\fs6$~, 
Dec = +30\arcdeg~47\arcmin\/~6\arcsec\/
and has a FWHM $\approx$ 1\arcsec\/. 
The photon statistics are too low to determine specific model parameters;
its spectrum is consistent with its being a 
bright feature inside the overall diffuse emission.

The source at RA = 01$^\mathrm{h}$~34$^\mathrm{m}$~$36\fs4$, 
Dec = +30\arcdeg~47\arcmin\/~16\arcsec\/ with a FWHM = 
1.08\arcsec\/, is most likely an AGN. For the spectrum, we get \nh\ =
$(6.3\pm0.9) \times 10^{22}$~cm$^{-2}$, 
with a photon index of $\Gamma = 2.3\pm0.4$.  
The flux (absorbed) $F_{\mathrm X}$(2-10~keV) = 3.2 
$\times 10^{-14}$~ergs~cm$^{-2}$~s$^{-1}$.
This source is harder than the emission from NGC\,604 and is nearly as 
bright as the diffuse emission.  The significantly higher value of the
\nh\/ argues that this is a background AGN, likely with a significant
amount of intrinsic absorption.

The SNR GKL98-94 is observed at RA = 01$^\mathrm{h}$~34$^\mathrm{m}$~33\fs0, 
Dec = +30\arcdeg~46\arcmin\/~40\arcsec\/, with an 
extent of 3.4$\pm$0.3\arcsec\/.  
This source is also faint with only \about100 counts detected.
It is well fit with a simple thermal model ({\tt APEC} in {\tt XSPEC})
with \nh\ = $0.56 (+0.08,-0.18) 
\times 10^{22}$~cm$^{-2}$, $kT = 0.25 (+0.08,-0.02)$~keV, and $F_{\mathrm 
X}$(0.5-2~keV) = 2.5 $\times 10^{-15}$~ergs~cm$^{-2}$~s$^{-1}$. 
According to
these fits, both the SNR and the diffuse emission in NGC\,604 show
excess absorption compared to the Galactic column (\nh\/ is
roughly an order of magnitude higher than expected) which is likely
due to significant amounts of dust and gas local to \m\/, associated
with the \hii\/ region.

\subsubsection{The Southern Spiral Arm Region}\label{sec-southarm}

\begin{figure*}
\centering
\epsscale{0.6}
\plotone{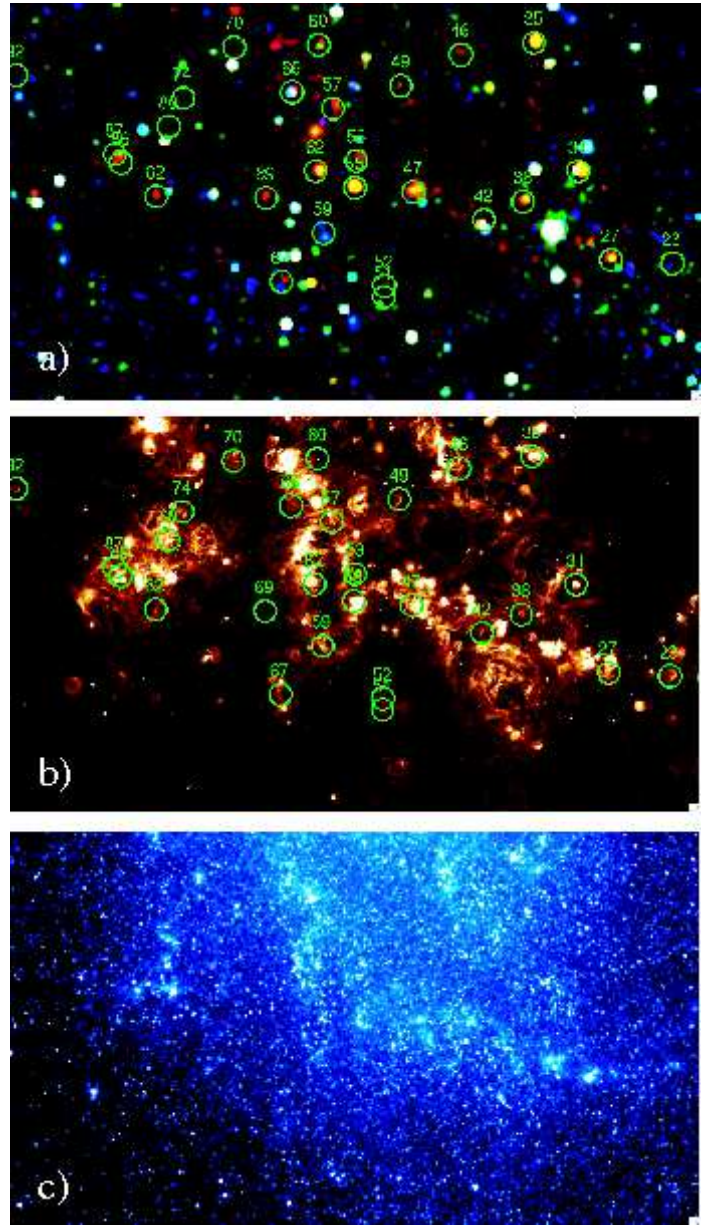}
\caption{\label{southarm}
A 9.3\arcmin\/ by 15.6\arcmin\/  region in the southern spiral arm of 
\m\ in three different wave bands. Panel a) shows a three-color image 
of the \chase\/ data, with the soft band shown in red, medium 
band in green, and hard band in blue. Panel b) shows the LGGS \halpha\/ image 
after subtraction of the stellar background. Panel c) shows the \spitzer\ 
IRAC 3.6~$\mu$m image of the region, which is dominated by stellar light. SNRs 
from the GKL98 optical SNR catalog are indicated in the top two panels with 
15\arcsec\/ diameter circles and ID numbers from that paper. Intensities are 
all shown on a log scale. Note the numerous coincidences of optical SNRs with 
soft \chandra\/ sources in the southern spiral arm region.
}
\end{figure*}

It is known that there are a large number of SNRs located in the
southern spiral arm of \m\/. 
Figure~\ref{southarm} shows a $9\farcm3\times15\farcm6$ region 
around this spiral arm in different energy bands.
In the first panel, we present the \chandra\/ three-color image
using the same bands in \S~\ref{sec-mosaics}, where
the second panel shows the continuum-subtracted 
\halpha\/ image from the Local Group Galaxies Survey 
\citep[LGGS,][]{massey2006}, and the third panel is the \spitzer\ 
IRAC 3.6~$\mu$m image which is dominated by stars.
The positions of the optical SNRs of the GKL98 list are
marked by the circles labeled with the GKL98 catalog number.
The images show that the optical SNRs are well-aligned with the southern 
spiral arm and that many of them show soft X-ray emission.
There are 26 SNRs from the GKL98 catalog contained in this
region,  10 of which have
counterparts in the \chase\/ source list (see Table~\ref{source_ids} for
details). Of the 16 SNRs without \chase\/ counterparts, many show
some evidence of subthreshold X-ray emission.
GKL98
-46,~-49,~-67,~-69, and~-82 seem similar in that there is some 
X-ray emission near the position of the optical SNR, but the emission 
was too faint to satisfy the selection criteria for this version of
our catalog.  These
types of sources may well become significant detections when all of
the \chase\/ data are combined.
Several other sources are worthy of note in this
image.  
GKL98-31 (\chase\/ \#123) is a bright X-ray source with a
spectrum dominated by emission lines of O, Ne, and Mg.  This SNR appears
to be an analog of the X-ray O-rich SNR in the Small Magellanic
Cloud 0103-72.6 \citep{park03} and will be the subject of a forthcoming
paper (Hughes~\etal\/ 2007, private communication). 
These data demonstrate the importance of \chandra's high
angular resolution for identifying the X-ray counterparts of the
optical SNRs in such a crowded region and for cleanly separating the
X-ray emission from one object from the surrounding objects for
spectral analysis.

\section{Summary}\label{sec-summary}

\begin{figure*}
\centering
\plotone{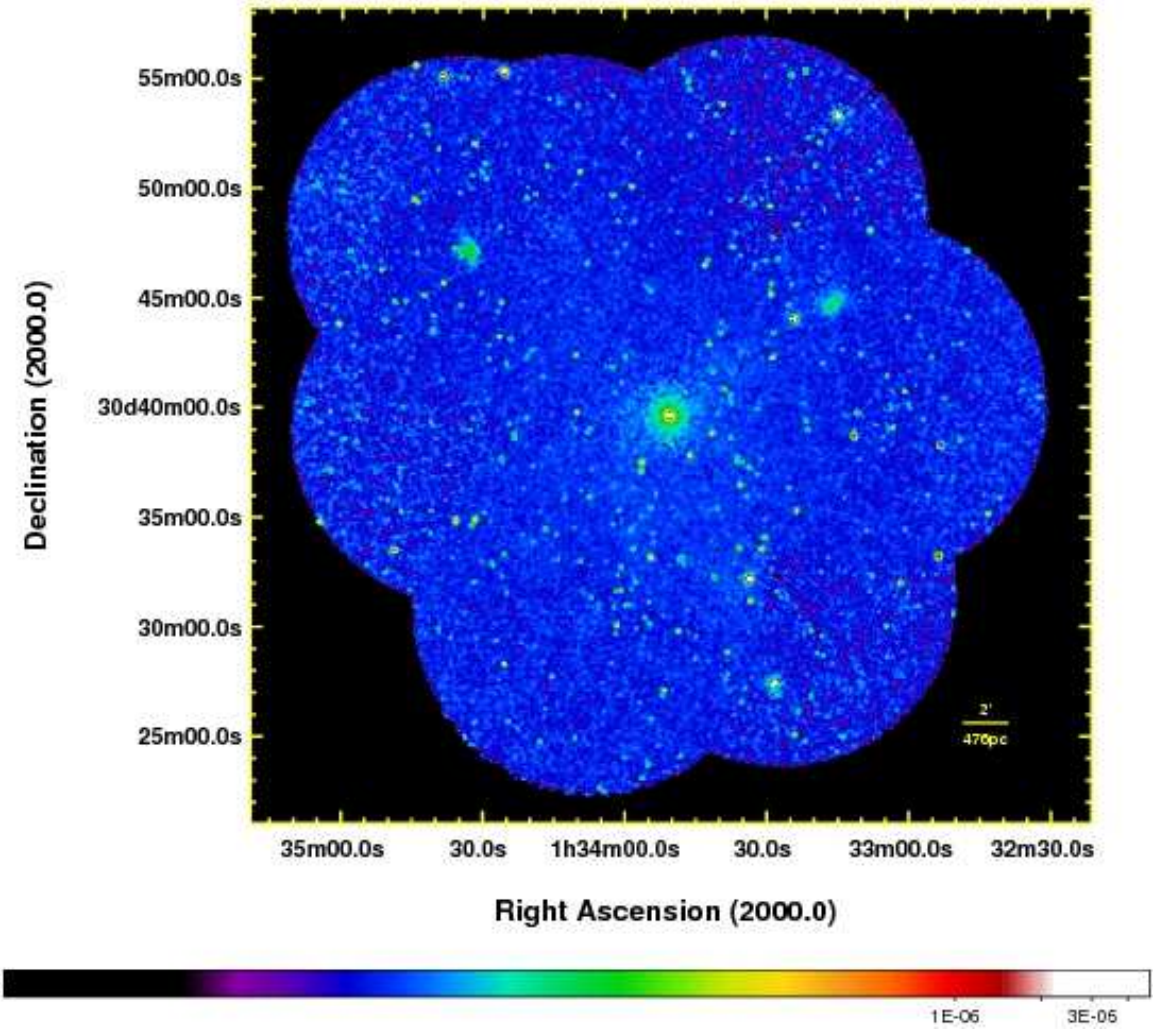}
\caption{\label{mosaic}
Exposure-corrected mosaic image of all \m\ pointings with ACIS-I as the primary
instrument (the \chase\/ observations and OBSIDs 1730 and 2023) in 
units of ${\rm ct~cm^{-2}~s^{-1}}$, in
the band 0.35 -- 8.0~keV.  Events within 8\arcmin\/ off-axis
are used.
}
\end{figure*}

\begin{figure*}
\centering
\plotone{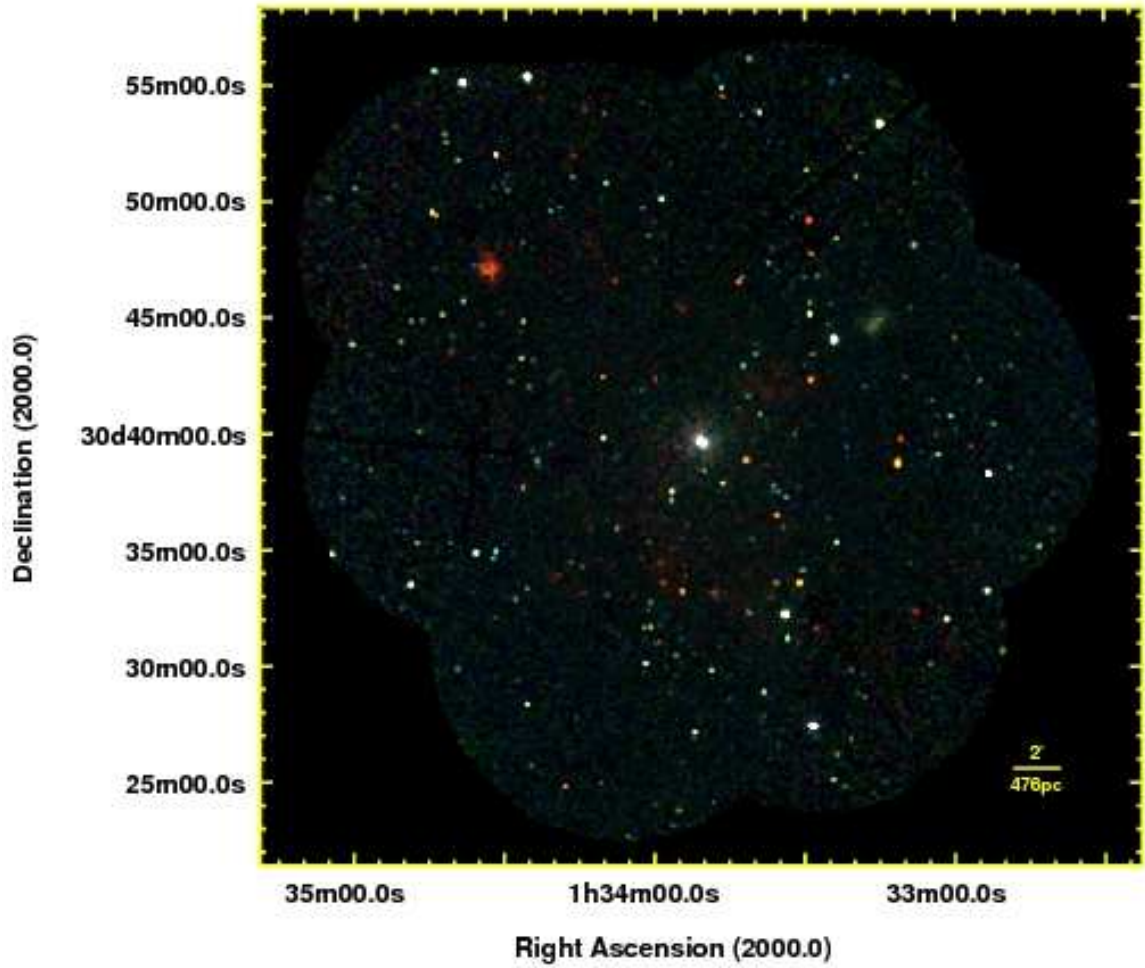}
\caption{\label{rgbmosaic}
Three-color mosaic image of all \m\ pointings with ACIS-I as the primary
instrument (the \chase\/ observations and OBSIDs 1730 and 2023), 
(red: 0.35 -- 1.1~keV, green: 
1.1 -- 2.6~keV, and blue: 2.6 -- 8.0~keV). Events are selected for off-axis 
angles $< 8$\arcmin\/. The images are exposure corrected. 
}
\end{figure*}

We have presented results from roughly the first half of the
\chase\/ survey of \m\/,  which covers the full area of the survey
but at typically half the depth.
Our  total intensity image in the 0.35--8.0~keV band 
(Fig.~\ref{mosaic}) is dominated by the
bright nucleus and surrounding diffuse emission, bright XRBs in \m\/
(such as X-7), bright diffuse objects such as NGC~604 and IC~131,
and bright background AGN and foreground stars. 
Our X-ray three-color
image (Fig.~\ref{rgbmosaic}) constructed from the three bands 
0.35--1.1~keV, 1.1--2.6~keV, and 2.6--8.0~keV,
distinguishes the sources with soft spectra such as the SNRs in \m\/
and NGC~604 from those with hard spectra such as the nucleus and X-7.
The first catalog from the survey includes 394
sources significant at the $3\sigma$ level as calculated from
the source and background counts determined by {\tt ACIS Extract}.  
We plotted the hardness ratios of the sources in the 
catalog in order to explore their spectral characteristics.  The SNRs, 
foreground stars, and \hii\/ regions separate from the harder XRBs and AGN 
in our plot. This information will be used in conjunction with optical 
followups to identify the sources as foreground or background source or 
sources within M33.

  Our cross-correlation with the previous X-ray catalogs from
\einstein\/~\citep{trinchieri88} and
\xmmnewton\/~(PMH04,MPH06) resulted in the overlap of
many sources. 
However, the agreement with the GMZ05 \chandra\/ catalog is not
as good, given the lower significance threshold employed by GMZ05. 
If we restrict our
cross-correlation to only include sources which are significant at the
$3\sigma$ level in the GMZ05 catalog, we find that there are 144 such
sources in the GMZ05 catalog and of these 144 sources, 112 have
a counterpart in the \chase\/ catalog within 3\arcsec\/ (increasing to
122 counterparts if the search radius is expanded to 10\arcsec\/). 
The unmatched sources will be prime candidates to study for variability.
We concluded from these cross-correlations that the agreement between the
\chase\/ catalog and the \einstein\/ and \xmmnewton\/ catalogs is
excellent and the agreement with the GMZ05 catalog is excellent also
provided that catalog is filtered to include only sources
significant at the $3\sigma$ level.  Given the new associations
discussed in this paper, the number of X-ray sources associated with
an optical and/or radio SNR has grown to 31 out of a possible 100
SNRs identified in the optical and the radio.

With a little over half of the data analyzed, \chase\/ is already
yielding exciting results by resolving confused regions and supernova 
remnants and by providing spectral information for source classification 
as well as lightcurves for exotic X-ray binaries and transient X-ray 
sources.  In addition to improving upon these measurements and adding to 
the source catalog, our ongoing analysis of the full data set will 
characterize the completeness of the sample, the contribution of 
background sources, and the large-scale diffuse emission from the ISM.

\acknowledgments

Support for this work was provided by the National Aeronautics and Space
Administration through \chandra{} Award Number G06-7073A issued by the 
\chandra{}
X-ray Observatory Center, which is operated by the Smithsonian
Astrophysical Observatory for and on behalf of the National Aeronautics
Space Administration under contract NAS8-03060. 
PPP, TJG, and RJE  acknowledge support under NASA contract NAS8-03060.
This work has made use of {\tt ACIS Extract}\footnote{\tt
http://www.astro.psu.edu/xray/docs/TARA/ae\_users\_guide.html},  SAOImage 
DS9\footnote{\tt http://hea-www.harvard.edu/RD/}, developed 
by the Smithsonian Astrophysical Observatory (Joye \& Mandel 2003), 
the {\tt XSPEC}\footnote{\tt http://xspec.gsfc.nasa.gov/}
spectral fitting package (Arnaud 1996), the 
FUNTOOLS\footnote{\tt http://hea-www.harvard.edu/RD/funtools} 
utilities package, the HEASARC 
FTOOLS\footnote{\tt 
http://heasarc.gsfc.nasa.gov/docs/software/lheasoft/ftools/}
package, and the 
CIAO\footnote{\tt http://cxc.harvard.edu/ciao/}
(\chandra{} Interactive Analysis of Observations) package.

{\it Facilities:} \facility{CXO (ACIS)}




\clearpage


\begin{deluxetable}{lcccccc}
\tabletypesize{\footnotesize}
\tablecaption{\label{obstab} List of \chase\ observations as of
1 September 2006 and archival observations used in this analysis.
}
\tablehead{
\colhead{Obs.} & 
\colhead{Field} & 
\multicolumn{2}{c}{Pointing direction} & 
\colhead{Start time (UT)} & 
\colhead{Roll angle} & 
\colhead{Exposure\tablenotemark{a}}\\
\colhead{ID} & 
\colhead{No.} & 
\colhead{RA} & 
\colhead{Dec} & 
\colhead{} & 
\colhead{[\degr]} &
\colhead{[ks]} \\
\colhead{} & 
\colhead{} & 
\multicolumn{2}{c}{(J2000.0)} & 
\colhead{} &
\colhead{} & 
\colhead{} }
\startdata
\multicolumn{3}{l}{ChASeM33 Observations} & & & & \\ 
\hline
6376 & 1 & 01:33:50.80 & +30:39:36.6 & 2006/03/03 21:15:13 & 308.48 & 94 \\
6378 & 2 & 01:34:13.80 & +30:47:48.1 & 2005/09/21 00:15:16 & 140.21 &104 \\
6380 & 3 & 01:33:33.90 & +30:48:40.7 & 2005/09/23 20:08:35 & 140.21 & 90 \\
6382 & 4 & 01:33:08.87 & +30:40:24.9 & 2005/11/23 13:02:17 & 262.21 & 72 \\
7226 & 4 & 01:33:08.87 & +30:40:24.9 & 2005/11/26 04:51:28 & 262.21 & 25 \\
6383 & 4 & 01:33:08.87 & +30:40:24.9 & 2006/06/15 06:23:36 & 97.80 & 91 \\
6384 & 5 & 01:33:27.90 & +30:31:25.3 & 2005/10/01 22:06:27 & 145.71 & 22 \\
7170 & 5 & 01:33:27.90 & +30:31:25.3 & 2005/09/26 06:36:36 & 145.71 & 41 \\
7171 & 5 & 01:33:27.90 & +30:31:25.3 & 2005/09/29 04:34:52 & 145.71 & 38 \\
6386 & 6 & 01:34:07.70 & +30:30:32.8 & 2005/10/31 09:18:33 & 224.21 & 15 \\
7196 & 6 & 01:34:07.70 & +30:30:32.8 & 2005/11/02 07:57:10 & 224.21 & 23 \\
7197 & 6 & 01:34:07.70 & +30:30:32.8 & 2005/11/03 01:52:16 & 224.21 & 13 \\
7198 & 6 & 01:34:07.70 & +30:30:32.8 & 2005/11/05 07:59:56 & 224.21 & 21 \\
7199 & 6 & 01:34:07.70 & +30:30:32.8 & 2005/11/06 02:09:08 & 224.21 & 15 \\
7208 & 6 & 01:34:07.70 & +30:30:32.8 & 2005/11/21 23:52:30 & 259.45 & 12 \\
6387 & 6 & 01:34:07.70 & +30:30:32.8 & 2006/06/26 01:48:05 & 103.21 & 77 \\
7344 & 6 & 01:34:07.70 & +30:30:32.8 & 2006/07/01 01:53:40 & 103.21 & 21 \\
6388 & 7 & 01:34:33.14 & +30:38:44.5 & 2006/06/09 22:58:32 & 94.87 & 89 \\
\hline
\multicolumn{3}{l}{Archival Observations} & & & & \\ 
\hline
1730 & 1 & 01:33:50.80 & +30:39:36.6 & 2001/07/12 08:29:46 & 108.58 & 49 \\
2023 &   & 01:34:32.90 & +30:47:04.0 & 2001/07/06 23:06:51 & 106.56 & 90 \\
\enddata
\tablenotetext{a}{Exposure time after screening out background flares.}
\end{deluxetable}

\begin{deluxetable}{rcccrrrrr}
\tablewidth{0pt}
\tabletypesize{\small}
\tablecaption{\label{newtab} ChASeM33 X-ray Sources}
\tablehead{
  \colhead{Source}
  &\colhead{RA (J2000)}
  &\colhead{DEC (J2000)}
  &\colhead{err ($''$)}
  &\colhead{Net\_Counts}
  & \multicolumn{1}{c}{Rate$^{a,e}$ $(\times 10^{-4})$} 
  & \multicolumn{1}{c}{Flux$^{b,e}$ $(\times 10^{-6})$} 
  & \multicolumn{1}{c}{HR1$^{c,e}$}
  & \multicolumn{1}{c}{HR2$^{d,e}$} \\
\colhead{} 
& \colhead{}             
& \colhead{}             
& \colhead{}            
& \colhead{$(\mathrm{ct})$}
& \multicolumn{1}{c}{($\mathrm{ct}\,\mathrm{s}^{-1}$)}
& \multicolumn{1}{c}{($\mathrm{photon}\,\mathrm{s}^{-1}\,\mathrm{cm}^2$)}
& \colhead{}  
& \colhead{} 
}
\startdata
1\phm{$^{E}$}& 01 32 24.53 & +30 33 22.7 & 0.67 & 49.8 & 5.1~~(1.1) & 4.0 (0.8) & $0.07$ (0.23) & $-0.02$ (0.18) \\
2\phm{$^{E}$}& 01 32 27.23 & +30 35 21.2 & 0.59 & 27.9 & 2.9~~(1.0) & 1.5 (0.5) & $-0.11$ (0.31) & $0.03$ (0.26) \\
3\phm{$^{E}$}& 01 32 29.20 & +30 36 18.1 & 0.36 & 99.3 & 10.2~~(1.3) & 5.1 (0.6) & $0.06$ (0.12) & $-0.04$ (0.10) \\
4\phm{$^{E}$}& 01 32 29.29 & +30 45 13.6 & 0.64 & 40.4 & 2.1~~(0.5) & 2.5 (0.6) & $-0.04$ (0.25) & $-0.21$ (0.19) \\
5\phm{$^{E}$}& 01 32 30.27 & +30 35 48.2 & 0.48 & 49.4 & 5.1~~(1.0) & 2.3 (0.4) & $0.10$ (0.18) & $0.09$ (0.17) \\
6\phm{$^{E}$}& 01 32 30.45 & +30 36 18.6 & 0.32 & 76.8 & 7.9~~(1.2) & 4.5 (0.6) & $-0.43$ (0.17) & $-0.37$ (0.09) \\
7\phm{$^{E}$}& 01 32 32.71 & +30 40 29.6 & 0.43 & 25.8 & 1.4~~(0.4) & 0.9 (0.3) & $0.15$ (0.30) & $0.03$ (0.29) \\
8\phm{$^{E}$}& 01 32 36.95 & +30 32 30.1 & 0.48 & 131.6 & 13.5~~(1.5) & 6.2 (0.6) & $0.16$ (0.10) & $0.03$ (0.10) \\
9\phm{$^{E}$}& 01 32 40.87 & +30 35 49.3 & 0.39 & 63.3 & 3.4~~(0.6) & 1.6 (0.2) & $0.01$ (0.18) & $-0.27$ (0.13) \\
10\phm{$^{E}$}& 01 32 41.35 & +30 32 18.3 & 0.52 & 118.3 & 4.1~~(0.5) & 2.2 (0.3) & $0.13$ (0.11) & $0.35$ (0.13) \\
\enddata
\tablecomments{The complete version of this table is in the electronic
edition of the Journal.  The printed edition contains only a sample of
the first ten rows.}
\tablenotetext{a}{Count Rate (0.35--8.0\,keV); error listed in parentheses}
\tablenotetext{b}{Absorbed photon flux (0.35--8.0\,keV); error listed in
parentheses. The photon flux is computed from the sum
of the photon fluxes in the 0.35-1.1, 1.1-2.6,and 2.6-8.0\,keV
bands. The photon fluxes are computed using the net
counts in the band, the mean effective area in the band,
and the exposure. The mean effective area for each band is
computed assuming a flat spectrum.}
\tablenotetext{c}{HR1 value, error listed in parentheses}
\tablenotetext{d}{HR2 value, error listed in parentheses}
\tablenotetext{e}{Pileup affects the derived values for Rate, Photon Flux,
            and HRs.  These values should be treated with caution for
            the bright sources.}
\tablenotetext{E}{Extended Source}
\end{deluxetable}

\begin{deluxetable}{ccccl}
\tablecaption{Catalog Correlations }
\tablehead{\colhead{ChASeM33} & 
 \colhead{XMM$^a$} & 
 \colhead{XMM$^b$} & 
 \colhead{Chandra$^c$} & 
 \colhead{Other~IDs/Comments$^d$} 
}
\tabletypesize{\scriptsize}
\tablewidth{0pt}\startdata
1 &  30 &  28 &  ~ &  ~ \\ 
3 &  37 &  36 &  ~ &  ~ \\ 
4 &  35 &  35 &  ~ &  ~ \\ 
6 &  39 &  ~ &  ~ &  ~Star(USNO1206-0019336(14.7);K3$^e$) \\ 
8 &  47 &  44 &  ~ &  ~Eclipsing~Binary$^f$ \\ 
9 &  56 &  51 &  ~ &  ~ \\ 
10 &  ~ &  ~ &  ~ &  ~Transient$^g$ \\ 
12 &  58 &  ~ &  ~ &  ~ \\ 
13 &  60 &  54 &  ~ &  ~ \\ 
15 &  64 &  57 &  ~ &  ~ \\ 
16 &  68 &  ~ &  ~ &  ~ \\ 
17 &  69 &  62 &  ~ &  ~ \\ 
21 &  70 &  64 &  ~ &  ~ \\ 
23 &  71 &  65 &  ~ &  ~ \\ 
24 &  73 &  67 &  ~ &  ~ \\ 
26 &  80 &  ~ &  ~ &  ~ \\ 
27 &  83 &  75 &  J013253.4+303817 &  ~M33X-1~ \\ 
28 &  85 &  76 &  J013253.9+303312 &  ~M33X-2~ \\ 
29 &  84 &  ~ &  ~ &  ~ \\ 
30 &  87 &  77 &  ~ &  ~ \\ 
33 &  91 &  81 &  ~ &  ~ \\ 
35 &  93 &  83 &  ~ &  ~SNR(GKL98-9)~ \\
\tablecomments{The complete version of this table is in the
electronic edition ofthe Journal.  The printed edition contains only a sample.}
\tablenotetext{a}{ Pietsch, et al.\ (2004)}
\tablenotetext{b}{ Misanovic et al.\ (2006)}
\tablenotetext{c}{ Grimm et al.\ (2005)}
\tablenotetext{d}{ SNRs from Gordon et al.\ (1998); UIT sources from Massey et al.\ (1996)}
\tablenotetext{e}{ Hatzidimitriou et al.\ (2006)}
\tablenotetext{f}{ Pietsch et al.\ (2006b)}
\tablenotetext{g}{ Williams et al.\ (2007)}
\enddata 
\label{source_ids}
\end{deluxetable}


\clearpage


\clearpage

\clearpage

\clearpage

\clearpage

\clearpage

\clearpage

\clearpage

\clearpage

\clearpage

\end{document}